\documentclass[twocolumn,english,aps,psfig,amssymb]{revtex4}
\usepackage[T1]{fontenc}
\usepackage[latin1]{inputenc}
\usepackage{float}
\usepackage{amsmath}
\usepackage{graphicx}
\usepackage{amssymb}

\makeatletter
\usepackage{amsmath}
\usepackage{epsfig}
\newcommand{\D}{\displaystyle}

\usepackage{babel}
\makeatother
\begin{document}

\title{Temperature-dependent quantum pair potentials and their application to dense partially ionized hydrogen plasmas}

\author{A.V.~Filinov$^{1}$, V.O.~Golubnychiy$^{2}$, M.~Bonitz$^{1}$,
 W.~Ebeling$^{3}$, and J.W.~Dufty$^{4}$}

\address{$^{1}$Institut f\"ur Theoretische Physik und Astrophysik,
Christian-Albrechts-Universit\"{a}t Kiel,
Leibnizstr. 15, D-24098 Kiel, Germany}

\address{$^{2}$Fachbereich Physik, Universit\"{a}t Rostock,
 Universit\"{a}tsplatz 3, D-18051 Rostock, Germany}

\address{$^{3}$Institut f\"{u}r Physik, Humboldt-Universit\"{a}t Berlin,
 Invalidenstrasse 110 D-10115 Berlin}

\address{$^{4}$University of Florida, Department of Physics,
PO Box 118440, Gainesville FL 32611-8440}

\begin{abstract}
Extending our previous work~\cite{filinov-etal.jpa03ik} we present a
detailed discussion of
accuracy and practical applications of finite-temperature
pseudopotentials for two-component Coulomb systems. Different
pseudopotentials are discussed: i) the diagonal Kelbg potential,
ii) the off-diagonal Kelbg potential iii) the {\em improved}
diagonal Kelbg potential, iv) an effective potential obtained with
the Feynman-Kleinert variational principle v) the ``exact'' quantum pair
potential derived from the two-particle density matrix. For the {\em
improved} diagonal Kelbg potential a simple temperature
dependent fit is derived which accurately reproduces the ``exact''
pair potential in the whole temperature range.
The derived pseudopotentials are then used in path integral Monte Carlo (PIMC) and
molecular dynamics (MD) simulations to obtain thermodynamical
properties of strongly coupled hydrogen. It is demonstrated that classical MD
simulations with spin-dependent interaction potentials for the electrons allow for an accurate description of the internal energy of hydrogen in the difficult
regime of partial ionization down to the temperatures of about $60
\, 000$ K.
Finally, we point out an interesting relation between the quantum
potentials and effective potentials used in density functional
theory.
\end{abstract}
\maketitle

\section{Introduction}

\label{intro}

In recent years there is growing interest in the properties of
dense {\em quantum} plasmas, particulary in astrophysics, laser
plasmas, and condensed matter, see
Refs.~\cite{boston97,binz96,green-book,bonitz-book,Haberland,kbt99}
for an overview. In particular, the thermodynamic properties of
hot dense plasmas are essential for the description of plasmas
generated by strong lasers \cite{Haberland}. Further, among the
phenomena of current interest are the high-pressure
compressibility of deuterium \cite{dasilva-etal.97}, metallization
of hydrogen \cite{weir-etal.96} and the hypothetical plasma phase
transition, e.g.
\cite{NormanStarostin,kremp,saumon,schlanges,BeEb99,fil_etal,red-book}
which occur in situations where both \emph{interaction and quantum
effects} are relevant.

While the case of strong degeneracy and the weak coupling limit have
been extensively studied theoretically, e.g. within the random phase
approximation,
plasma properties at {\em intermediate coupling and degeneracy} (when
 $\Gamma$ - the ratio of the potential energy to the mean kinetic energy
exceeds unity), are a hot topic of the
present research activity. For an overview of present day
analytical methods,  see e.g.
Refs.~\cite{boston97,binz96,bonitz-book,BeEb99}. Analytical
methods typically use a chemical picture where electrons, ions and
bound states (atoms, molecules etc.) are treated as
independent species, and the chemical composition (degree of
ionization) is computed from a mass action law (non-ideal Saha
equation).  However, these methods are based on perturbation expansions in
the coupling strength and are thus limited to regions of small
coupling parameters, $\Gamma<1$ or $r_{s}<1$ ($r_s$ is the quantum coupling
parameter, $r_s={\bar r}/a_B$). Furthermore, the
mass action law becomes increasingly inaccurate in the region
where the electrons are degenerate (uncertainty in the mass action
constants). Also, during rapid pressure ionization around the Mott
density the distinction between free and bound particles is an
open problem.

On the other hand, in the last decade, static properties (e.g.
equation of state) of dense hydrogen in thermal equilibrium
have been successfully investigated with ``exact''
quantum-statistical methods, such as path integral Monte Carlo
(PIMC)~\cite{Margo,Militzer,trigger, filinov2003}. This first
principle numerical technique is well suited for an accurate
treatment of many-particle correlation effects in quantum systems,
but unfortunately does not give dynamical characteristics of the plasma
(with an exception of those obtained within the linear response
theory). The alternative numerical approach for dense partially
ionized plasmas (which does not have the above shortcoming) is a
group of methods based upon ab initio quasi-classical molecular
dynamic simulations (MD), e.g. Refs.~\cite{vova01,KTR94}, when a
real quantum system is projected onto a classical one where most of
quantum effects are included in some effective interparticle
interaction potentials, such as the ones proposed by Kelbg~\cite{Ke63},
Deutsch~\cite{Deu}, Klakow, Toepffer and Reinhard~\cite{KTR94} and
many others, e.g.~\cite{Rogers,Deu2,perrot}. These potentials can
be derived from the two-particle Slater sum using Morita's method.

However, no rigorous comparison of the accuracy of these
potentials has been done yet, which is one of the aims of this
paper. Different quantum potentials are compared with an ``exact''
pair potential obtained from the two-particle density matrix.
Furthermore, we introduce pair potentials including particle
statistics, e.g. describing interaction between electrons in the
singlet and triplet states, and use them in our MD simulations of
two-component hydrogen plasmas.

This paper is organized as follows: In Section~\ref{ep_s} we
discuss different methods to obtain an effective quantum pair
potential. In the weak coupling limit, this potential leads
exactly to the off-diagonal Kelbg potential the properties of
which are being discussed and compared to its commonly used
diagonal approximation. We outline two methods for a direct
solution of the off-diagonal two-body Bloch equation which are
then used in Section~\ref{comp_pot} for numerical comparison with
the Kelbg and improved Kelbg potentials for rigorous assessment of
the accuracy of the latter. Further, in Section~\ref{pimc_results}
we present an analysis of the accuracy of the diagonal and off-diagonal
Kelbg potentials in the PIMC simulations. Section \ref{md_results}
describes an application of the improved Kelbg potentials to
classical molecular dynamics simulations of dense hydrogen.
Comparing the results to those of PIMC simulations allows us to
conclude that use of the improved Kelbg potential allows to
significantly extend the range of applicability of classical MD to
the region of partial ionization and to temperatures as low as
approximately one third of the binding energy. Section~\ref{dft}
discusses another field of potential applicability of the quantum
potentials -- density functional theory. Finally,
Section~\ref{dis} concludes the paper.

\section{Effective quantum pair potentials} \label{ep_s}

In this section we discuss different possibilities to obtain
effective quantum potentials describing interactions in the
two-particle problem.

\subsection{Analytical solution of two-body Bloch equation. Off-diagonal and diagonal Kelbg
potential}\label{kelbg_eq}

The equilibrium pair density matrix at a given inverse
temperature $\beta=1/k_BT$
is the solution of the two-particle Bloch equation

\begin{align}
\frac{\partial}{\partial \beta} \,
& \rho ({\textbf{r}}_{i}, {\textbf{r}}_{j},
{\textbf{r}}'_{i}, {\textbf{r}}'_{j}; \beta)
 = - \hat {H} \, \rho ({\textbf{r}}_{i}, {\textbf{r}}_{j},
 {\textbf{r}}'_{i}, {\textbf{r}}'_{j}; \beta),
 \nonumber \\
& \hat {H} = \hat {K}_{i} + \hat {K}_{j} +
\hat {U}({\textbf{r}}_{i}, {\textbf{r}}_{j},
{\textbf{r}}'_{i}, {\textbf{r}}'_{j}).
\label{bloch}
\end{align}
Numerical methods to obtain the density matrix of the
Eq.~(\ref{bloch}) will be considered in Sec.~\ref{be_s}. Here, we
concentrate on the available analytical solutions in the limit of
weak coupling. If the interaction is weak, the Eq.~(\ref{bloch})
can be solved by perturbation theory with the following
representation for the two-particle density matrix
\begin{eqnarray}
\rho_{ij}= &  & \frac{(m_{i}m_{j})^{3/2}}{(2\pi\hbar\beta)^{3}}\exp\left[-\frac{m_{i}}{2\hbar^{2}\beta}({\textbf{r}}_{i}-{\textbf{r}}_{j}')^{2}\right]\nonumber \\
 &  & \times\exp\left[-\frac{m_{j}}{2\hbar^{2}\beta}({\textbf{r}}_{i}-{\textbf{r}}_{j}')^{2}\right]\exp[-\beta\Phi_{ij}],\label{full_dm_Kelbg}\end{eqnarray}
 where $i,j$ are particle indices, $\rho_{ij}\equiv\rho({\textbf{r}}_{i},{\textbf{r}}_{j},{\textbf{r}}'_{i},{\textbf{r}}'_{j};\beta)$,
and $\Phi_{ij}\equiv
\Phi({\textbf{r}}_{i},{\textbf{r}}_{j},{\textbf{r}}_{i}',{\textbf{r}}_{j}';\beta)$
is the off-diagonal two-particle effective potential. In the
following we will consider application of this result to Coulomb
systems. As a result of first-order perturbation theory we get
explicitly

\begin{eqnarray}
\Phi^0({\textbf{r}}_{ij},{\textbf{r}}'_{ij},\beta) & \equiv &
q_{i}q_{j}
\int_{0}^{1}\frac{d\alpha}{d_{ij}(\alpha)}{\textrm{erf}}
\left(\frac{d_{ij}(\alpha)/\lambda_{ij}}{2\,\sqrt{\alpha(1-\alpha)}}\right),
\nonumber \\
\label{kelbg-off}
\end{eqnarray}
where
$d_{ij}(\alpha)=|\alpha{\textbf{r}}_{ij}+(1-\alpha){\textbf{r}}'_{ij}|$,
${\textrm{erf}}(x)$ is the error function,
${\textrm{erf}}(x)=\frac{2}{\sqrt{\pi}}\int_{0}^{x}dte^{-t^{2}}$,
and $\lambda_{ij}^{2}=\frac{\hbar^{2}\beta}{2\mu_{ij}}$ with
$\mu_{ij}^{-1}=m_{i}^{-1}+m_{j}^{-1}$. The diagonal element
(${\textbf{r}}'_{ij}={\textbf{r}}_{ij}$) of~(\ref{kelbg-off}) is
the potential derived by Kelbg and his
coworkers~\cite{Ke63,green-book}
\begin{eqnarray}
\Phi^0(x_{ij})=\frac{q_{i}q_{j}}{\lambda_{ij}x_{ij}}\,
\left\{1-e^{-x_{ij}^{2}}+\sqrt{\pi}x_{ij}
\left[1-{\textrm{erf}}(x_{ij})\right]\right\} \label{kelbg-d}
\end{eqnarray}
 with $x_{ij}=|{\textbf{r}}_{ij}|/\lambda_{ij}$. The Kelbg potential is finite at zero distance
reflecting that it captures the basic quantum diffraction effects
and the quantum nature of two-particle interaction at small
distances which prevents any divergence. From Eq.~(\ref{kelbg-d})
it is also clear that quantum effects become dominant (and there
the quantum potential deviates from the classical Coulomb
potential) at distances $r_{ij}\lesssim \lambda_{ij}$ given by the
thermal DeBroglie wavelength. We will see below that, in
interacting systems, this is only a rough approximation, and at
strong coupling, the expression for the quantum particle
``extension'' deviates strongly from $\lambda_{ij}$ and needs to
be generalized.

To obtain a simplified expression for the rather complex
 quantum potential
(\ref{kelbg-off}) one can approximate the off-diagonal
matrix elements by
the diagonal ones. A first possibility is to approximate the integral
over $\alpha$ by the length of the interval multiplied with the integrand
in the center (Mittelwertsatz) which leads to the so-called KTR-potential
due to Klakow, Toepffer and Reinhard which (in the diagonal approximation)
is often used in quasi-classical MD simulations \cite{KTR94}
\begin{eqnarray}
\Phi^0({\textbf{r}}_{ij},{\textbf{r}}'_{ij},\beta)
\equiv\frac{q_{i}q_{j}}{d_{ij}(1/2)}\,{\textrm{erf}}
\left(\frac{d_{ij}(1/2)}{\lambda_{ij}}\right),\label{KTR-pot}
\end{eqnarray}
 where $d_{ij}(1/2)=\frac{1}{2}|{\textbf{r}}_{ij}+{\textbf{r}}'_{ij}|$.
Alternatively, the integral can be simplified by taking the off-diagonal
Kelbg potential only at the center coordinate,
\begin{equation}
\Phi^0({\textbf{r}}_{ij},{\textbf{r}}'_{ij},\beta)
\approx\Phi_{ij}^0\left(\frac{|{\textbf{r}}_{ij}|+|{\textbf{r}}'_{ij}|}{2},
\beta\right).
\end{equation}

Many authors use the \emph{end-point}
approximation~(\ref{kelbg-d}) for the effective potential
$\Phi({\textbf{r}}_{ij},{\textbf{r}}'_{ij},\beta)$ in the pair
density matrix~(\ref{full_dm_Kelbg}) due to the fact that it is
very convenient computationally. The pair potential for
interparticle interaction simply is replaced by an effective
potential which has only a dependence on the radial
variables~$|{\textbf{r}}_{ij}|,\,|{\textbf{r}}'_{ij}|$. However,
most of the accuracy is usually lost in this end-point
approximation.

Since the Kelbg potential is obtained by first order perturbation
theory its application is limited to weak coupling,
$\Gamma\lesssim1$, where $\Gamma$ is the ratio of mean potential
to kinetic energy. In unbound and bound states of an
electron-proton pair this results in the following conditions on
temperature
\begin{eqnarray}
\Gamma &=& \frac{e^{2}}{\bar{r}}/k_{B}T\lesssim1\hspace{0.2cm}
\Rightarrow\hspace{0.2cm}k_{B}T\gtrsim\frac{e^{2}}{\bar{r}}\nonumber \\
\Gamma &=& \text{Ry}/k_{B}T\lesssim1\hspace{0.2cm}
\Rightarrow\hspace{0.2cm}k_{B}T\gtrsim \text{Ry},
\end{eqnarray}
where $\text{Ry}=\text{Ha}/2=e^2/2 a_B$, and $a_B$ is the Bohr
radius. For the last case the Kelbg potential (and any of the
simplifying approximations) can be only valid for temperatures
sufficiently above the atomic binding energy, i.e for the case of
hydrogen, $T\gtrsim \text{Ry}/k_B \approx158\;000$~K. We address
this point in more detail in the section~\ref{pimc_results}, where
``exact'' binding energies and pair correlation functions for an
electron-proton pair are compared with the results obtained with
the potentials~(\ref{kelbg-off}) and (\ref{kelbg-d}).

\subsection{Improved diagonal Kelbg potential}\label{improv_s}

The limitation of the Kelbg potential to describe quantum systems
only when there are no bound states has lead several
researchers~\cite{gombert,wagenknecht01,filinov-etal.jpa03ik} to
introduce and investigate a more generalized form of the quantum
potential with an additional free parameter $\gamma_{ij}$
\begin{eqnarray}
\label{impr_kelbg} \Phi\left({r}_{{ij}},\beta\right) & =&
\\ &  & \hspace{-1.5cm}=\frac{q_{i}q_{j}}{r_{ij}}\,
\left[1-e^{-\frac{r_{ij}^{2}}{\lambda_{ij}^{2}}}+
\sqrt{\pi}\frac{r_{ij}}{\lambda_{ij}\gamma_{ij}}
\left(1-{\textrm{erf}}\left[\gamma_{ij}\frac{r_{ij}}{\lambda_{ij}}\right]\right)\right].
\nonumber
\end{eqnarray}
This potential has an advantage of preserving the correct first
derivative at $r=0$ of the original Kelbg potential,
$\Phi(0,\,\beta)^{\prime}_{r}=-q_{i}q_{j}/\lambda_{ij}^{2}$, but
at the same time allows to correct the wrong value of the height of the Kelbg
potential at $r=0$, i.e.
$\Phi(0,\,\beta)=q_{i}q_{j}\sqrt{\pi}/(\lambda_{ij}\,\gamma_{ij})$
to include bound states. Using the definition of the effective
potential as
 \begin{equation}
e^{-\beta \Phi_{ij}}\equiv S_{ij},
\end{equation}
where $S_{ij}\equiv S(r_{ij},\beta)$ is the exact binary Slater sum
of particles $i,j$. The fit parameter
$\gamma_{ij}$ in Eq.~(\ref{impr_kelbg}), is related to the Slater sum at zero interparticle distance according to
\begin{equation}
\gamma_{ij} =-\frac{\sqrt{\pi}}{\lambda_{ij}}
\frac{q_{i}q_{j}\,\beta}{{\rm ln}[S_{ij}(r_{ij}=0,\beta)]}.
\label{gamma_ij}
\end{equation}
It is important to note that $\gamma_{ij}$ depends both on
temperature and the type of particles. For example, the binary
Slater sum of two electrons at zero separation has the form
(including the average ``$\langle \ldots \rangle$'' over possible
values of the total spin $S=0,1$)
\begin{eqnarray}
S_{ee}^{\langle \rangle}(r_{ee}=0,\beta)=2\sqrt{\pi}\xi_{ee}J_{1}(\xi_{ee}), \nonumber \\
J_{1}(\xi_{ij})=\int\limits _{0}^{\infty}e^{-x^{2}}\frac{x\,
dx}{1-\exp(-\frac{\pi\xi_{ij}}{x})}, \label{s_ee_avr}
\end{eqnarray}
where the interaction parameter
$\xi_{ij}=q_{i}q_{j}\,\beta/\lambda_{ij}$.

On the other hand, for an electron-proton pair the Slater sum can be written as
\begin{eqnarray}
S_{ep}(r_{ep}=0,\beta) & = & 4\sqrt{\pi}\xi_{ep}J_{1}(\xi_{ep})+
\sqrt{\pi}\xi_{ep}^{3}Z_{3}(\xi_{ep}), \nonumber \\
Z_{n}(\xi) & = & \sum_{y=1}^{\infty}y^{-n}e^{\xi^{2}/4y^{2}},
\label{s_ep}
\end{eqnarray}
where the last term shows the contribution of the bound states.

The original Kelbg potential was derived for very high
temperatures without taking into account exchange between
particles. This work was followed by several studies where the
pseudopotentials for identical particles have been calculated
numerically~\cite{storer, barker} or
analytically~\cite{Davis,rohde,gombert2} using expansions in a
quantum parameter, small particle separation, and temperature. In the
present work, following these studies, we approach the problem
of the pseudopotential with exchange by using the formalism of
two-particle density matrices (DM). The pair DM can be calculated
numerically (see Sec.~\ref{be_s}) or expressed in analytical
form~(\ref{full_dm_Kelbg}) using the improved Kelbg
potential~(\ref{impr_kelbg}).

In the case of a pair of electrons, they can be in a singlet or
triplet state, and the spatial wave function is symmetric or
antisymmetric under the exchange of particle indices. Thus, one
can define a binary effective electron-electron interaction for
three different cases
\begin{eqnarray}
e^{-\beta U^{S(T)}_{ij}}&=&\frac{\rho^{[2]}({\textbf{r}}_{i},
{\textbf{r}}_{j}, {\textbf{r}}_{i}, {\textbf{r}}_{j}; \beta)\pm
\rho^{[2]}({\textbf{r}}_{i}, {\textbf{r}}_{j}, {\textbf{r}}_{j},
{\textbf{r}}_{i}; \beta)}{\rho^{[1]}({\textbf{r}}_{i},
{\textbf{r}}_{i};\beta)\rho^{[1]}({\textbf{r}}_{j},
{\textbf{r}}_{j})}, \nonumber \\
e^{-\beta U_{ij}^{\langle \rangle}}&=&\frac{3}{4} e^{-\beta
U^{T}_{ij}} + \frac{1}{4} e^{-\beta U^{S}_{ij}},
\end{eqnarray}
where $\rho^{[1]}$ and $\rho^{[2]}$ are the one- and two-particle
density matrices, and $U^{S}_{ij}$, $U^{T}_{ij}$ and
$U_{ij}^{\langle \rangle}$ are the effective interactions in the
singlet (S), triplet (T) state and the spin-averaged potential,
respectively.

If we now approximate the two-particle DM, $\rho^{[2]}$, by
Eq.~(\ref{full_dm_Kelbg}) and factorize it into the DM's of the
center-of-mass and relative coordinates (the corresponding
expressions are given in Sec.~\ref{be_s}, cf.
Eq.~(\ref{exact_diag_u})), then we obtain for the pseudopotential
between two electrons being in the singlet (triplet) state and
for the spin-averaged potential, respectively,
\begin{eqnarray}
U_{ee}^{S(T)} =-\frac{1}{\beta}{\rm ln}\left(e^{-\beta
U_{ee}({\textbf{r}},{\textbf{r}})} \pm e^{-
r^{2}/\lambda^2_{ee}}\, e^{-\beta
U_{ee}({\textbf{r}},-{\textbf{r}})}\right) \label{sing_t} \\
U_{ee}^{\langle \rangle}=-\frac{1}{\beta}{\rm ln}\left(e^{-\beta
U_{ee}({\textbf{r}},{\textbf{r}})} -\frac{1}{2} e^{-
r^{2}/\lambda^2_{ee}}\, e^{-\beta
U_{ee}({\textbf{r}},-{\textbf{r}})}\right) \label{u_e}
\end{eqnarray}
In this expression, the function,
$U_{ee}({\textbf{r}},{\textbf{r}^{\prime}})$, is a pseudopential between
distinguishable particles (i.e. calculated without exchange effects).
Thus, for it one can substitute the original Kelbg
potential, Eq.~(\ref{kelbg-d}), the improved Kelbg potential, Eq.~(\ref{impr_kelbg}),
or any further improved approximation for the binary interaction. In the
case of two electrons, if for $U_{ee}({\textbf{r}},{\textbf{r}})$
we use the improved Kelbg potential~(\ref{impr_kelbg}), then the
fit parameter $\gamma_{ee}$ must be obtained from
Eq.~(\ref{gamma_ij}) where for the binary Slater sum one should take
the one for two distinguishable particles with Coulomb repulsion
\begin{equation}
S_{ee}^{no\,
exc}(r_{ee}=0,\beta)=4\sqrt{\pi}\xi_{ee}J_{1}(\xi_{ee}).
\label{noexc_s}
\end{equation}

As it follows from Eq.~(\ref{u_e}) an exchange contribution
(effect of particle statistics in the pair interaction) arises
from the kinetic energy part of the density matrix and the
non-diagonal potential, $U_{ee}({\textbf{r}},-{\textbf{r}})$,
which in the first order of the perturbation theory can be
calculated using Eq.~(\ref{kelbg-off}). A further simplification
(which is crucial for application of the pseudopotentials in
semiclassical MD simulations presented in Sec.~\ref{md_results})
can be achieved by approximating the off-diagonal potential by the
diagonal terms, $U_{ee}({\textbf{r}},-{\textbf{r}})\approx
\frac{1}{2}\left[U_{ee}({\textbf{r}},{\textbf{r}})+U_{ee}(-{\textbf{r}},-{\textbf{r}})\right]=
U_{ee}({\textbf{r}},{\textbf{r}})$, and the above expressions are
reduced to
\begin{eqnarray}
U_{ee, 0}^{S(T)}= U_{ee}({\textbf{r}},{\textbf{r}})-
\frac{1}{\beta} {\rm ln}\left\{1\pm e^{-
r^{2}/\lambda^2_{ee}}\right\}, \label{u_e_d} \\
U_{ee, 0}^{\langle \rangle}= U_{ee}({\textbf{r}},{\textbf{r}})-
\frac{1}{\beta} {\rm ln}\left\{1-\frac{1}{2} e^{-
r^{2}/\lambda^2_{ee}}\right\}.
\end{eqnarray}
We can note that in the diagonal approximation for the potential,
the exchange term corresponds to the case of the ideal Fermi gas
(i.e. exchange without interaction), the exchange term arising
from the interaction is missing.

Taking in Eq.~(\ref{u_e_d}) the limit $r\rightarrow 0$ we see that
the potential of the triplet state shows a logarithmic divergency
\begin{equation}
U_{ee, 0}^T= U_{ee}({\textbf{r}},{\textbf{r}})-2k_B
T\ln\left\{\frac{r}{\lambda_{ee}}\right\}+O\left(\frac{r^2}{\lambda^2_{ee}}\right),
\end{equation}
whereas the singlet and the spin-averaged potential acquires an
additional exchange contribution
\begin{eqnarray}
U_{ee, 0}^{S,\langle \rangle}=U_{ee}({\textbf{r}},{\textbf{r}})\mp
k_B T\,{\rm ln}\left\{2\right\} +
O\left(\frac{r^2}{\lambda^2_{ee}}\right),
\end{eqnarray}
but the slope of these potentials at the origin are same as in the
case without exchange. This means, in case of Coulomb interaction,
the slope is defined by the slope of the original Kelbg potential
$\Phi^0$,
\begin{equation}
U_{ee}({\textbf{r}},{\textbf{r}})|_{r\rightarrow 0}=
\Phi^0_{ee}(0)-\frac{e^2
r}{\lambda^2_{ee}}+O\left(\frac{r^2}{\lambda^2_{ee}}\right).
\end{equation}

In our previous paper~\cite{filinov-etal.jpa03ik} we reported on
the temperature dependence of the fitting parameter $\gamma_{ij}$
for the electron-electron and electron-proton interactions. There,
two types of calculations have been presented. First, the values
of $\gamma(\beta)$ were obtained by a least-square fit of the
improved diagonal Kelbg potential (IDKP), the
Eq.~(\ref{impr_kelbg}), to the ``exact'' pair potential $U$~(see
Eq.~(\ref{exact_diag_u})), and second, from Eq.~(\ref{gamma_ij})
by evaluating the values of the binary Slater sums. It has been
found that both methods agree within the statistical uncertainty.

\begin{figure}
\hspace{-1.0cm}
\includegraphics[width=7.0cm,angle=-90]{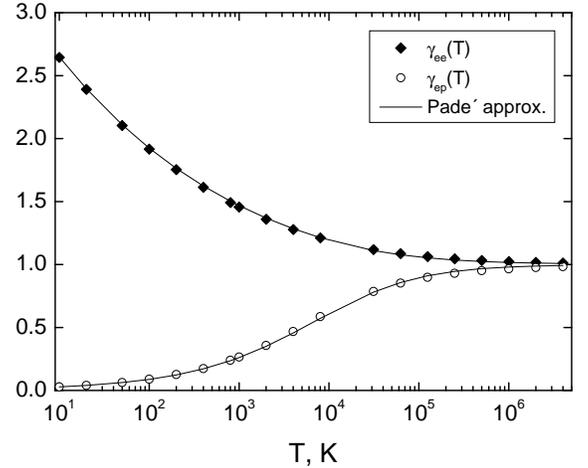}
\caption{Temperature dependence of the fit parameter for the
binary interactions: electron-proton, $\gamma_{ep}(T)$, and
electron-electron (no exchange), $\gamma_{ee}(T)$.
Symbols show the $\gamma$-values obtained with the least-square
fit of the IDKP to the ``exact'' pair potential without exchange.
Solid curves correspond to the Pad\'e
approximation~(\ref{ep_fit_gamma},\ref{ee_up_down_fit_gamma}).}
\label{gamma_fig}
\end{figure}

Extending our earlier results, we now present a Pad\'e
approximation which contains an analytical temperature dependence
of the parameters $\gamma_{ij}$ which will be useful for practical
applications,
\begin{eqnarray}
\gamma_{ep}(T)
 & = & \frac{x_{1}+x_{1}^{2}}{1+a_{ep}\, x_{1}+
 x_{1}^{2}},
 \label{ep_fit_gamma}\\
\gamma_{ee}(T) & = & \frac{\gamma_{ee}(T\rightarrow0)+ a_{ee}\, x_{1}+x_{1}^{2}}{1+x_{1}^{2}},
\label{ee_up_down_fit_gamma}
\end{eqnarray}
 where $x_{1}=\sqrt{8\pi k_BT/{\rm Ha}}$ \,
(with the Hartree energy, ${\rm Ha}=2{\rm Ry}=315\;775$~K),
$a_{ep}=1.090(14)$, $a_{ee}=0.18(1)$. The limit value,
$\gamma_{ee}(T\rightarrow0)$, has been obtained from
Eq.~(\ref{gamma_ij}) by evaluating the zero temperature limit of
the binary Slater sum~(\ref{noexc_s})
\begin{equation}
\gamma_{ee}(T\rightarrow0)
\approx-\frac{2}{\sqrt{\pi}}\tilde{x}^{3}\frac{1}{\ln\{8\tilde{x}^{4}/
\sqrt{\pi}\}-3\tilde{x}^{2}},
\end{equation}
 with $\tilde{x}=\left(|\pi\,\xi_{ee}|/2\right)^{\frac{1}{3}}$.
 The excellent accuracy of the Pad\'e approximation is demonstrated in
 Fig.~\ref{gamma_fig}.

In Fig.~(\ref{gamma_fig}) we present the temperature dependence of
$\gamma$ obtained from the least-square fit (full and open
symbols) to the ``exact'' pair potential of distinguishable
particles (no exchange) and Pad\'e
approximation~(\ref{ep_fit_gamma}) (solid curves). The most
important result is that the corrected Kelbg potential is now {\em
not limited to weak coupling} as is the original Kelbg potential.
For the case $\gamma_{ij}=1$, Eq.~(\ref{impr_kelbg}) coincides
with Eq.~(\ref{kelbg-d}). One clearly sees the deviation of
$\gamma_{ij}$ from unity for $T \leq 10^6$ K,  which shows that
the quantum extension of particles is becoming influenced by
interaction effects and is now of the order of $\tilde
\lambda_{ij}=\lambda_{ij}\gamma_{ij}$, instead of the original
thermal DeBroglie wavelength $\lambda_{ij}$. Thus, with the Pad\'e
formulas~(\ref{ep_fit_gamma}-\ref{ee_up_down_fit_gamma}), we have
obtained an analytical fit for the quantum extension of the
scattering particles.

The Pad\'e approximations~(\ref{ep_fit_gamma}), (\ref{ee_up_down_fit_gamma})
have been  successfully used in quasi-classical molecular dynamics
simulations of two-component hydrogen plasmas.
As we show in Sec.~\ref{md_results}, they allow to obtain accurate
results for partially hydrogen (or other quantum
systems of oppositely charged particles with bound states).

Finally, we want to note that, in the MD simulations of quantum
plasmas, it can be advantageous to have spin-dependent
potentials for the electron subsystem defined by
Eqs.~(\ref{sing_t}) or ~(\ref{u_e_d}). The spin-resolved
approximation allows for a refined modelling, for example, it
allows for the description of molecule formation, spin density
waves, spin flip processes in the presence of a magnetic field and so on.

\subsection{Effective potentials from numerical solution of two-body Bloch equation}
\label{be_s}
In the present section, we briefly describe the numerical methods
which have been used to solve the two-particle problem in order to
obtain the ``exact'' quantum pair potentials. These results have been
used to obtain the analytical fit in the improved Kelbg potential.
Furthermore, they will be used to test the accuracy of various analytical
approximations for the quantum pair potentials in Sec.~\ref{comp_pot} below.

Let us factorize full two-particle density matrix into a
center-of-mass term and a density matrix of relative coordinates
\begin{equation}
\rho({\textbf{r}}_{i},{\textbf{r}}_{j},
{\textbf{r}}_{i}',{\textbf{r}}_{j}';\beta) = \rho_{{\rm
cm}}(\mathbf{R},{\bf \mathbf{R}}';\beta)
\rho(\mathbf{r},\mathbf{r}';\beta), \label{full_dm}
\end{equation}
where
${\textbf{R}}=(m_{i}{\bf{r}}_{i}+m_{j}{\bf{r}}_{j})/(m_{i}+m_{j})$,
and ${\textbf{r}}={\textbf{r}}_{i}-{\textbf{r}}_{j}$, and
analogously for ${\textbf{R}}',\,{\textbf{r}}'$. When for the
relative DM in the analogy with the Eq.~(\ref{full_dm_Kelbg}) we
can define the effective pair potential as
\begin{equation}
\rho(\mathbf{r},\mathbf{r}';\beta) =
\rho_{kin}(\mathbf{r},\mathbf{r}';\beta)\, e^{-\beta
U(\mathbf{r},\mathbf{r}';\beta)},
\end{equation}
which results in the following expression
\begin{equation}
 U(\mathbf{r},\mathbf{r}';\beta)=-\frac{1}{\beta}
 \ln\left[\rho(\mathbf{r},\mathbf{r}';\beta)/\rho_{kin}(\mathbf{r},\mathbf{r}';\beta)\right].
 \label{exact_diag_u}
 \end{equation}
where $\rho(\mathbf{r},\mathbf{r}';\beta)_{kin}$ is the kinetic
energy DM.

One of the possibilities to get the relative density matrix
$\rho(\mathbf{r},\mathbf{r}';\beta)$ is to directly solve the
corresponding one-particle Schr\"odinger equation and calculate
the DM as a contribution from bound and continuum states. This
procedure is advantageous when the Schr\"odinger equation can be
solved analytically and we know analytical expressions for
contributions of scattering and bound states, as for example for
Coulomb potential, e.g.~\cite{gombert2}. But if it is not the
case, a separate calculation of each matrix element for each new
values of end-points $\textbf{r}$ and $\textbf{r}^{\prime}$ will
be not efficient and time-demanding procedure. In principle, such
calculations can be done in advance with
$U(\textbf{r},\textbf{r}^{\prime};\beta)$ be stored in the tables
of the potential, but one still needs to solve the Schr\"odinger
equation many times for each value of quantum numbers and also for
wavefunctions of continuum states.

It is possible to approach to this problem from the other side and
calculate the DM directly without solving the Schr\"odinger
equation. In this work we apply two efficient methods -- the
\emph{matrix squaring technique}~\cite{storer,ceperley95rmp} and
the Feynman-Kleinert \emph{variational
approach}~\cite{Kleinert,Kleinert2}. In the Sec.~\ref{comp_pot} we
will compare an accuracy of the pseudopotnetials obtained with
these methods.

\subsubsection{Matrix squaring technique} \label{matsq}

The exact off-diagonal pair density matrix can be calculated
efficiently by this method introduced by Storer and
Klemm~\cite{storer}. For the case of spherical symmetry of the
interaction potential, the relative pair density matrix in the
Eq.~(\ref{full_dm}) is expanded in terms of partial waves. This
expansion reads, for the two- and three-dimensional cases,
\begin{eqnarray}
\rho^{2D}({\textbf{r}},{\textbf{r}}';\beta)
& = & \frac{1}{2\pi\sqrt{r\, r'}}\sum_{l=-\infty}^{+\infty}\;
\rho_{l}(r,r';\beta)e^{i\, l\Theta},\label{part_wave2d}\\
\rho^{3D}({\textbf{r}},{\textbf{r}}';\beta) & = & \frac{1}{4\pi
r\, r'}\sum_{l=0}^{+\infty}\;(2l+1)\; \rho_{l}(r,r';\beta)\;
P_{l}(\cos\Theta),\nonumber \end{eqnarray} where $\Theta$ is the
angle between ${\textbf{r}}$ and ${\textbf{r}}'$. Each
partial-wave component satisfies the $1$D Bloch equation for a
single particle in an external potential given by the interaction
potential and also a convolution equation,
\begin{equation}
\rho_{l}(r,r';\tau)=\D\int\limits _{0}^{\infty}dr''\;
\rho_{l}(r,r'';\tau/2)\;\rho_{l}(r'',r';\tau/2).
\label{basic_square}
\end{equation}
 This is the basic equation of the \emph{matrix-squaring method}
 which allows to calculate the function $\rho_l$ at a given temperature
 $1/\tau$ from the same function at a two times higher temperature.
Squaring the density matrix $k$ times results in a lowering
of the temperature by a factor of $2^{k}$. Each squaring involves
only a one-dimensional integration which, due to the Gaussian-like
nature of the integrand in Eq.~(\ref{basic_square}), can be performed
quite accurately and efficiently by standard numerical procedures.
To start the matrix-squaring iterations, Eq.~(\ref{basic_square}),
one needs a known accurate high-temperature form for the density matrix.
A convenient choice is the semiclassical approximation
\begin{equation}
\rho_{l}(r,r';\tau)=\rho_{l}^{0}(r,r';\tau)
\exp{\left(-\frac{\tau}{|r-r'|}\D\int_{r}^{r'}V(x)dx\right)},
\end{equation}
 where $\rho_{l}^{0}(r,r';\tau)$ is the partial-wave component of
the free-particle density matrix.

Once the pair density matrix $\rho_{l}(r,r';\tau)$ is computed for
the desired value of $\tau$, it is substituted into
Eqs.~(\ref{part_wave2d}-\ref{basic_square}),
and a summation over partial waves readily yields the full relative
density matrix.

\subsubsection{Variational perturbation approach}
\label{var_app}

As a second method for solving the off-diagonal Bloch equation we
used a \emph{variational perturbation expansion} developed by Feynman
and Kleinert~\cite{Kleinert}. In this procedure the initial density
matrix is presented in the form of a trial path integral which consists
of a suitable superposition of local harmonic oscillator path integrals
centered at arbitrary average positions ${\textbf{x}}_{m}$, each
with its own frequency squared $\Omega^{2}({\textbf{x}}_{m})$. One starts
from decomposing the action in the density matrix as
\begin{eqnarray}
\rho({\textbf{r}},{\textbf{r}}';\beta)
& = & \int\limits _{({\textbf{r}},0)\rightarrow({\textbf{r}}',\hbar\beta)}
{\mathcal{D}}\,{\textbf{x}}\; e^{-A[{\textbf{x}}]/\hbar},
\label{action1}
\\
A[{\textbf{x}}]
& = & A_{\Omega,\,{\textbf{x}}_{m}}[{\textbf{x}}]+A_{int}[{\textbf{x}}],
\label{action2}
\end{eqnarray}
 with $A_{\Omega,\,{\textbf{x}}_{m}}[{\textbf{x}}]$ being the action
of a trial harmonic oscillator with the potential minimum located
at ${\textbf{x}}_{m}$, and ${\cal D}$ being the functional integral
over all trajectories. The interaction part
\begin{equation}
A_{int}[{\textbf{x}}]=\int_{0}^{\hbar\beta}d\eta
\left[V\left[{\textbf{x}}(\eta)\right]
-\frac{1}{2}\mu\;\Omega^{2}\;[{\textbf{x}}(\eta)
-{\textbf{x}}_{m}]^{2}\right],
\label{action_int}
\end{equation}
 is defined as the difference between the original potential $V({\textbf{x}})$
and the displaced harmonic oscillator. The $\Omega^{2}$ term in
Eq.~(\ref{action_int})
compensates the contribution of $A_{\Omega,\,{\textbf{x}}_{m}}[{\textbf{x}}]$
in Eq.~(\ref{action2}). Now one can calculate the density matrix~(\ref{action1})
by treating the interaction~(\ref{action_int}) as a perturbation,
leading to a moment expansion
\begin{multline}
\rho(\mathbf{r},\mathbf{r}';\beta)
=\rho_{0}^{\Omega,\,{\bf{x}}_{m}}({\bf{r}},{\bf{r}}';\beta)
\bigg(1-\frac{1}{\hbar} \langle
A_{int}[\mathbf{x}]\rangle_{\bf{r},\bf{r}'}^{\Omega,\mathbf{x}_{m}}
+\\
+\frac{1}{2\hbar^{2}}
\langle A_{int}^{2}[{\bf x}]\rangle_{\bf{r},\bf{r}'}^{\Omega,{\bf x}_{m}}-...
\bigg)
=e^{-\tau\, W_{N}^{\Omega,{\bf x}_{m}}}\;
\left(\frac{\mu}{2\pi\hbar^{2}\tau}\right)^{d/2},
\label{pertrub}\end{multline}
with the definition

\begin{eqnarray}
W_{N}^{\Omega,\mathbf{x}_{m}}
& = &
\frac{d}{2\beta}\ln\frac{\sinh\hbar\beta\Omega}{\hbar\beta\Omega}
+\frac{\mu\Omega}{2\hbar\beta\sinh\hbar\beta\Omega}\times
\nonumber \\
 &  & \times\left[(\widetilde{\mathbf{r}}^{2}+\widetilde{\mathbf{r}}'^{2})
 \cosh\hbar\beta\Omega-2\widetilde{\mathbf{r}\,}\widetilde{\mathbf{r}}'\right]
-\nonumber \\
 &  & -\frac{1}{\beta}\sum_{n=1}^{N}\frac{(-1)^{n}}{n!\hbar^{n}}
 \langle A_{int}[\mathbf{x}]\rangle_{\mathbf{r},\mathbf{r}'}^{\Omega,\mathbf{x}_{m}},
 \label{eff_pot}\end{eqnarray}
where~$d$ is the space dimensionality and $N$ the order of the
approximation. The function
$\rho_{0}^{\Omega,\,{\textbf{x}}_{m}}({\textbf{r}},{\textbf{r}}')$
is the trial harmonic oscillator density matrix,
$\widetilde{{\textbf{r}}}=({\textbf{r}}-{\textbf{x}}_{m}),\,
\widetilde{{\textbf{r}}}'=({\textbf{r}}'-{\textbf{x}}_{m})$, and
the expectation value of the interaction action on the r.h.s. of
Eq.~(\ref{eff_pot}) is given by
\begin{eqnarray}
&&
\langle A_{int}^{n}[{\textbf{x}}] \rangle_{{\textbf{r}},{\textbf{r}}'}^{\Omega,{\textbf{x}}_{m}}
=\frac{1}{\rho_{0}^{\Omega,\,{\textbf{x}}_{m}}({\textbf{r}},{\textbf{r}}')}
\D\int\limits _{\widetilde{{\textbf{r}}},0}^{\widetilde{{\textbf{r}}}',
\hbar\beta}{\mathcal{D}}\widetilde{{\textbf{x}}}\,
\prod_{l=1}^{n}
\bigg\{\D\int\limits _{0}^{\hbar\beta}d\tau_{l}\times
\nonumber \\
&&\qquad \times
V_{int}\left[\widetilde{{\textbf{x}}}(\tau_{l})+{\textbf{x}}_{m}\right]\,
e^{\{-\frac{1}{\hbar}A_{\Omega,\,{\textbf{x}}_{m}}
[\widetilde{{\textbf{x}}}+{\textbf{x}}_{m}]\}}
\bigg\}.
\end{eqnarray}
 The function $W_{N}^{\Omega,\,{\textbf{x}}_{m}}$ can be identified
as an \emph{effective quantum potential} which is to be optimized
with respect to the variational parameters
$\{\Omega^{2}({\textbf{r}},{\textbf{r}}';\beta),\,{\textbf{x}}_{m}({\textbf{r}},{\textbf{r}}';\beta)\}$.
Note that, in the high temperature limit, this effective potential
goes over to the original potential $V({\textbf{r}})$. The optimal
parameter values are determined from the extremity conditions
\begin{align}
\frac{\partial W_{N}^{\Omega, \,{\textbf{x}}_{m}}({\textbf{r}},
{\textbf{r}}';\beta)} {\partial \Omega^{2}} = 0, \; \;
\frac{\partial W_{N}^{\Omega, \,{\textbf{x}}_{m}}({\textbf{r}},
{\textbf{r}}';\beta)} {\partial {\textbf{x}}_{m}} = 0.
\end{align}
The perturbation series~(\ref{eff_pot}) is rapidly converging, in
most cases already in the first-order approximation
$W_{1}^{\Omega,\,{\textbf{x}}_{m}}$ for the effective potential,
and gives a reasonable estimate of the desired quantities.

\section{Comparison of the pair potentials and their temperature
dependence} \label{comp_pot}

We will now compare accuracy of the pair potentials discussed
above (or two-particle density matrices corresponding to these
potentials), their temperature dependence and range of
applicability.

\subsection{Full density matrix of electron-proton pair}

\begin{figure}
\includegraphics[
  width=9cm]{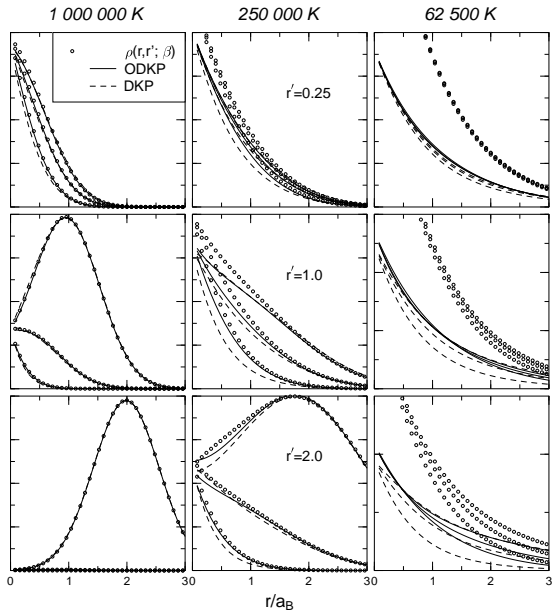}
\vspace{-.5cm} \caption{The ``exact'' off-diagonal density matrix
$\rho(r,r';\phi)$ for an electron-proton pair vs. the density
matrix calculated with the diagonal (DKP) and off-diagonal (ODKP)
Kelbg potentials. In all figures, results for three angular values
are given $\phi=0$ (upper curves), $\phi=\pi/2$ (middle) and
$\phi=\pi$ (lower curves). The proton is located at the origin,
and the vector ${\textbf{r}}'$ (initial electron position) is
fixed, $|{\textbf{r}}'|=0.25;\,1.0;\,2.0$. The vector
${\textbf{r}}$ (final electron position) is varied, $\phi$ is the
angle between the vectors \textbf{r} and \textbf{r}'.}

\label{ang_matrix}
\end{figure}

In Fig.~\ref{ang_matrix} we show the angular dependence of the
full off-diagonal two-particle density matrix calculated with the
off-diagonal Kelbg potential -- ODKP~(\ref{kelbg-off}) and its
diagonal approximation -- DKP~(\ref{kelbg-d}). The density matrix
is shown at several temperature values ($T=1\;000\;000,\;250\;000$
and $62\;500$ K) and several angular distances
($\phi=0,\,\pi/2,\,\pi$) between the vectors
${\textbf{r}}\equiv{\textbf{r}}_{ij},\,{\textbf{r}}'\equiv{\textbf{r}}'_{ij}$
(in each of the figures, the top curves correspond to the case of
parallel vectors, $\phi=0$, the lowest curves to antiparallel
vectors, $\phi=\pi$). Also, for reference, we give the
off-diagonal density matrix obtained from the `exact' solution of
the Bloch equation, cf. Sec.~\ref{matsq}. At high temperatures,
$T\geq 250\;000$ K, the Kelbg density matrix does not exhibit
large deviations from the exact result. At $T=1\;000\;000$K, the
ODKP density matrix practically coincides with the exact solution,
whereas the DKP approximation shows small deviations. In these
cases the perturbation expansion applies, $\Gamma \sim 0.15$, see
left column of Fig.~\ref{ang_matrix}. With decreasing temperature,
the deviations from the exact results grow, see middle column. To
better understand the details of the deviations, we magnified them
by including also results for $T=62\;500$ K, which is far beyond
the scope of the perturbation theory, $T \approx 0.4\;
\text{Ry}/k_B$, i.e. $\Gamma\approx 2.5$. Here we observe that, at
the origin, the density matrix of the Kelbg potential is $3$ times
less than the exact one. The largest errors were found for the
DKP, in particular, in the case when the vectors
${\textbf{r}},{\textbf{r}}'$ have the opposite direction
($\phi=\pi$).

%\vspace{-2.5cm} \hspace{-1.15cm}
\label{pair_dis}

The behavior of the full density matrix can be understood from the
following considerations. The density matrix results from
contributions of kinetic and potential energy operators, cf.
Eq.~(\ref{bloch}). At small distances ($r'=0.25$) the Coulomb
attraction between an electron and a proton dominates and,
therefore, the density matrix shows an exponential decay. At the
largest distance ($r'=2.0$) kinetic and potential energy are of
the same order and a Gaussian-like free particle density matrix
emerges, as can be clearly seen in the bottom left part of
Fig.~\ref{ang_matrix}.

From this first comparison we can conclude that both, the DKP and
the ODKP, show satisfactory agreement with the exact result in the
cases where perturbation theory applies,  $T\gtrsim2\, $Ry. At
lower temperatures there is only qualitative agreement. The
strongest deviations arise for small interparticle distances $\{
{\textbf{r}}$, ${\textbf{r}}' \}$, and this, as will be shown
below, results from the incorrect height of the Kelbg potential at
zero distance~$r=0$.

\subsection{Effective interaction of electron-proton and electron-electron
pairs} \label{pair_pot}

\begin{figure}[h]
\hspace{-0.9cm}
\includegraphics[height=9.2cm,angle=-90]{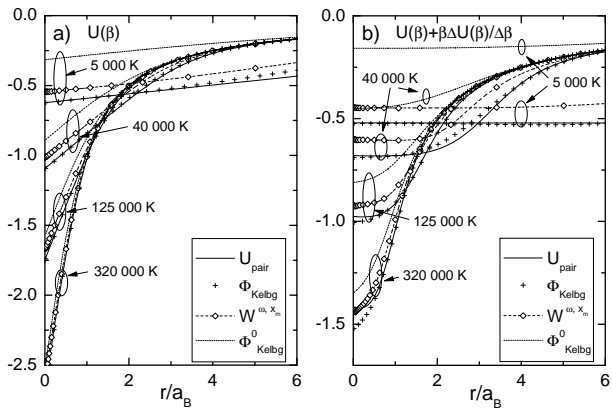}
\vspace{-0.50cm} \caption{{(\bf a):} Diagonal effective electron-proton
potential (in units of Ha) for several cases: the DKP
$\Phi^0({\textbf{r}};\beta)$~(\ref{kelbg-d}), the improved DKP
$\Phi({\textbf{r}};\beta)$~(\ref{impr_kelbg}), variational
potential $W_{1}^{\Omega,{\textbf{x}}_{m}}$~(\ref{eff_pot}), pair
potential $U_{p}$~(\ref{exact_diag_u}) corresponding to the
``exact'' density matrix. Each potential is given at at three
temperature values: $5\,000, 40\, 000, 125\, 000$ and $320\,
000$~K. {\bf (b):} the function
$U(\beta)+\beta\partial U(\beta)/\partial \beta$ for the same
approximations and temperatures. } \label{diag_eff_pot}
\end{figure}

In Fig.~\ref{diag_eff_pot}(a,b) we show and compare the accuracy
of several effective electron-proton potentials and their
temperature derivatives obtained by various methods. As an
``exact'' reference potential to which the accuracy of other
potentials is compared we use $U_{pair}$ obtained from the
electron-proton pair density matrix calculated with the matrix
squaring technique.

First, we note from Fig.~\ref{diag_eff_pot}(a) that, at given
temperatures $T\leq 2 \text{Ry}$, the original Kelbg potential
shows the largest deviations from the ``exact'' result,
$U_{pair}$. While the spatial derivative of the DKP coincides with
that of $U_{pair}$, a systematic offset
of the DKP compared to $U_{pair}$ is observed at the origin $r=0$,
which increases when the temperature is lowered. The agreement is
satisfactory only for the curve corresponding to $T= 320\, 000$~K.
The accuracy of the Kelbg potential becomes worse for quantities
involving its temperature derivative. For example, for
the total energy one has to compute the thermodynamic average of the
function  $\partial (\beta \, U(\beta)/\partial
\beta)|_{U=\Phi^0_{Kelbg}}$. This function is shown in Fig.~\ref{diag_eff_pot}(b).
If multiplied by the Boltzmann factor $e^{-\beta \,
U(\beta) }$, this function is a good estimate for the binding energy
($E_b$) of an electron-proton pair. In the case of a bound state
the main contribution to the energy comes from the region of small interparticle
distance, $r \lesssim 3 a_B$. Therefore, the behavior
of $\partial (\beta \, U(\beta)/\partial \beta)$ near the origin
determines the accuracy of the calculations of the energy and other
thermodynamic quantities. As we can see from the curve for $5\,000K$ in
Fig.~\ref{diag_eff_pot}(b), the depth of the DKP is much less than
that of $U_{pair}$  and, therefore, it
gives a too low binding energy of $E_b\approx 0.16$~Ha, i.e. a factor
of three too low compared with the true ground state energy $E^{0}_b=0.5$~ Ha.

As it was already discussed in Sec.~\ref{improv_s} the accuracy of
the DKP can be improved with the additional fit parameter $\gamma_{ij}$. In
Fig.~\ref{diag_eff_pot}(a,b) this potential is denoted as
$\Phi_{Kelbg}$. One can see that at all considered temperatures
$\Phi_{Kelbg}$ practically coincides with $U_{pair}$. Even in the
case of the strong coupling ($T=5\, 000$~K) the agreement is very
good.

The next potential showh in this figure is the variational potential,
$W^{\omega,\, x_m}$, introduced in the Sec.~\ref{var_app}. This
potential is more accurate than the DKP and qualitatively
reproduces the ``exact'' effective pair potential $U_{pair}$ for
temperatures $T=125\,000$~K, $320\, 000$~K and its derivative (see
Fig.~\ref{diag_eff_pot}(b)). The key point is that the
\emph{variational perturbation theory}~\cite{Kleinert} replaces
the perturbation expansion in $\Gamma$ (which does not
converge for $\Gamma\gtrsim 1$) into another expansion,
Eq.~(\ref{pertrub}), which does not have this restriction. The
results of this approach can be improved by taking into account
higher order terms in Eq.~(\ref{pertrub}) (the results shown in
the figure include only the first term, $n=1$). The convergence of
Eq.~(\ref{pertrub}) extends even to very strong coupling and has
been successfully applied in field theory~\cite{Kleinert2}.

%\vspace{-0.5cm}
%\hspace{-0.7cm}

We mention that comparison with other effective potentials has
been performed in our paper~\cite{filinov-etal.jpa03ik}. In particular,
the ``exact'' pair potential was compared with the results of
Barker~\cite{barker} (the calculations of the pair potential by
the direct eigenfunction summation) and the Deutsch potential. A
good agreement has been found with the data of~\cite{barker},
while the deviations of the Deutsch potential turned out to be
slightly larger than that of the Kelbg potential. The reason is that the Deutsch
potential has an incorrect spatial derivative, $U^{\prime}_r$, for
$r\lesssim 3 a_B$.

%\vspace{-2.7cm}
\begin{figure}
\hspace{-1.2cm}
\includegraphics[height=9.5cm,angle=-90]{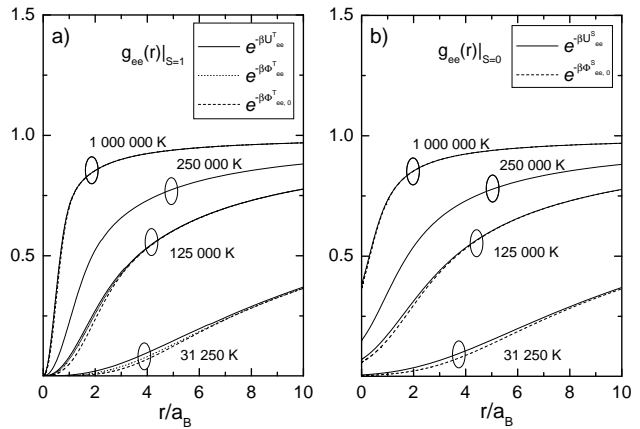}%
\vspace{-0.5cm} \caption{Pair distribution function for a pair of
electrons. \emph{a}: In a triplet state (parallel spins, $S=1$);
\emph{b}: In a singlet state (antiparallel spins, $S=0$). Cases
compared are: Eq.~(\ref{sing_t}), using for $U_{ee}$ the ``exact'' pair
potential and the Kelbg potential, $\Phi_{ee, 0}$, respectively,
as well as the simpler expression, Eq.~(\ref{u_e_d}) with the Kelbg
potential. The pair distribution functions are shown for four
temperatures, $T=31\,250$, $125\,000$, $250\,000$ and
$1\,000\,000$~K.} \label{ee_eff_pot}
\end{figure}

Next, in Fig.~\ref{ee_eff_pot}, we compare pair distribution
functions (PDF) of two electrons in a singlet and triplet states
for different temperatures, obtained from the expression with the
effective potential
\begin{equation}
g(r)\propto e^{-\beta U^{S(T)}(r)}.
\end{equation}
Due to the Pauli principle in Fig.~\ref{ee_eff_pot}(a) the PDF is
goes to zero as $r_{ee} \rightarrow 0$. On the other hand, for electrons in a
singlet state (Fig.~b) this happens only if the temperature is decreased
down to $31\; 250$~K, when the potential energy dominates over the
kinetic energy. The three lines in Fig.~\ref{ee_eff_pot}(a) show
the cases when as an effective potential in the
Eqs.~(\ref{u_e}-\ref{u_e_d}) we substitute the ``exact'' pair
potential and the Kelbg potential. In the last case,
Eq.~(\ref{u_e_d}), the exchange contribution from the potential
function is neglected. This, as shows Fig.~\ref{ee_eff_pot}(a),
becomes important only for the temperature $31 \; 250$~K and below.
But the overall accuracy of the Kelbg potential for a description of two
particles with the same charge (even without improving its value
at the origin with $\gamma$) is significantly better (compared to
the results with the ``exact'' pair potential) than for
particles with opposite charge, cf. Fig.~\ref{diag_eff_pot}.
This is due to absence, in this case, of contributions from bound
states.

The pair correlation functions at $T=125\; 000$~K can be compared
with those for a hydrogen plasma obtained by molecular dynamics simulations, see
Fig.~\ref{cap:Radial-distribution-functions}. One can see that, at
least at this temperature, the electron-electron PDF's of that
many-particle simulation look quite similar to the two-particle case shown in
Fig.~\ref{ee_eff_pot}. Some differences are noticeable for the value of
$g_{ee}$ at $r=0$ as the density is increased from
$r_s=6$ to $r_s=4$, see Sec.~\ref{md_results}.

In the next section we discuss application of quantum pair
potentials in thermodynamical calculations using Feynman
trajectories in imaginary time (PIMC).

\section{Quantum pair potentials in Path integral Monte Carlo} \label{pimc_results}

It is well known (see for example discussion in Chapter $12$
of~\cite{Kleinert}) that the singularity of the attractive Coulomb
potential causes difficulties in the euclidian path integral. If
based on Feynman's original path integral representation, a path
consists of a \emph{finite} number of straight pieces, each with a
classical euclidian action, containing the singular Coulomb
potential. However, in this case, the energy of the path can be
lowered indefinitely by an almost stretched configuration which
corresponds to a slowly moving particle sliding down to the
$-e^{2}/r$ abyss. This phenomenon is called \emph{path collapse}.

One possibility to prevent this effect is to use a modified
``regularized'' Coulomb potential which has a cutoff at $r=0$.
This procedure, however, is quite arbitrary, and the results are
sensitive to the used cutoff parameters. Of course, in nature,
these difficulties are prevented by quantum fluctuations which
equip the path with a configurational entropy. The latter must be
sufficiently singular to produce a regular free energy bounded
from below. The inclusion of quantum fluctuations in the Euclidean
action of the Feynman path pieces smoothes the singular Coulomb
potential, producing an effective potential that is finite at the
origin, and the \emph{path collapse} is avoided. This again shows
the importance of effective potentials, specifically, in
``quasiclassical'' simulations (classical Monte Carlo and
molecular dynamics methods). Of great importance are potentials
which have a closed analytical form. In this case for many
thermodynamical quantities it is possible to obtain analytical
solutions.

For simulations of correlated quantum many-body systems which are based on
\emph{first principles},
the initial many-body hamiltonian with the true singular Coulomb energy
operator is to be considered and solved to find some effective many-body
interaction potential. For this it is important that in the high-temperature
limit the N-particle density matrix can be expanded in terms of 2-particle,
3-particle etc. contributions. If the temperature is sufficiently
high then all contributions except the first one, taking into account
two-particle correlations, can be omitted. As a result, the following
approximation for the N-particle density matrix holds
\begin{eqnarray}
\label{pair_approx}
\rho(\mathbf{R},\mathbf{R}';\tau)
& \approx & \prod_{i}^{N}
\rho^{[1]}(\mathbf{r}_{i},\,{\mathbf{r}}'_{{i}};\tau)\times
\\\nonumber
& \times &\prod_{j<k}\frac{\rho^{[2]}(\mathbf{{r}}_{j},{\mathbf{r}}_{k},
 {\mathbf{r}}_{j}',{\mathbf{r}}_{k}';\tau)}
 {\rho^{[1]}(\mathbf{{r}}_{i},{\mathbf{r}}_{i}';\tau)
 \rho^{[1]}(\mathbf{{r}}_{k},{\mathbf{r}}_{k}';\tau)}+O(\rho^{[3]}),
\end{eqnarray}
where ${\textbf{R}}=\{{\textbf{r}}_{1},\dots,{\textbf{r}}_{N}\}$
specifies coordinates of all $N$ particles,
$\rho^{[1]}\;(\rho^{[2]})$ is the single (two) particle density
matrix. The above \emph{pair} approximation is usually used in
PIMC simulations~\cite{ceperley95rmp}. The N-particle density
matrix $\rho(\beta)$ contains a complete information about the
system with the observables given by
\begin{equation}
\langle\hat{O}\rangle=\frac{{{\rm Tr}
\left[\hat{O}\,\hat{\rho}(\beta)\right]}}
{{{\rm Tr}\left[\hat{\rho}(\beta)\right]}}
=\frac{\D\int d\mathbf{R}\,\langle\mathbf{R}|\hat{O}\,\hat{\rho}(\beta)|\mathbf{R}\rangle}
{\D\int d\mathbf{R}\,\langle\mathbf{R}|\hat{\rho}(\beta)|\mathbf{R}\rangle}.
\label{pimc_mean}
\end{equation}
Due to the exponential form, the N-particle density operator
$\hat{\rho}(\beta)=e^{-\beta\hat{H}}$ can be factorized (in
analogy to the matrix squaring method above) as
$\hat{\rho}(\beta)=\left[\hat{\rho}(\tau)\right]^{M}$ with
$M=\beta/\tau$. Consequently, the N-particle density operator
$\hat{\rho}(\beta)$ is expressed in terms of density operators at
an $M$ times higher temperature $1/\tau=M\cdot k_{B}T$. If $M$ is
chosen sufficiently large then one can apply the pair
approximation~(\ref{pair_approx}). Thus, accurate results for the
quantum pair potentials and, consequently, the pair density
matrix, will allow to compute the density matrix of the whole
$N-$particle system. Here we are not interested in the
investigation of the accuracy of approximation~(\ref{pair_approx})
but concentrate on the two-body problem where
Eq.~(\ref{pair_approx}) is exact.

It is clear that the observables~(\ref{pimc_mean}) computed with
the approximate pair-density matrix $\rho^{[2]}$ contain an error
of the order~$O(1/M^{2})$. Below we will investigate the
convergence, as a function of $M$, of main thermodynamic
properties (total energy and e-p pair distribution) for an
electron-proton pair using for the pair density matrix
$\rho^{[2]}$ results computed with the off-diagonal and the
diagonal Kelbg potential.

\subsection{Comparison of diagonal and off-diagonal Kelbg potential on the example of Hydrogen atom}
\label{off_dm_results}

We consider a hydrogen atom in a box with periodic boundary
conditions (box size, $L=20\, a_{B}$) at several temperatures,
$T=31\,250-62\,500$ K, when the hydrogen atom can ionize into free
particles, as well as for the case $T<10\,000$ K, when there is
essentially only the contribution from the atomic ground state.
First, in Fig.~\ref{prot_elect_f} we show the e-p pair
distribution functions (normalized to the volume~$dV=4\pi
r^{2}dr$).

\vspace{-0.7cm}
\includegraphics[%
  width=8cm]{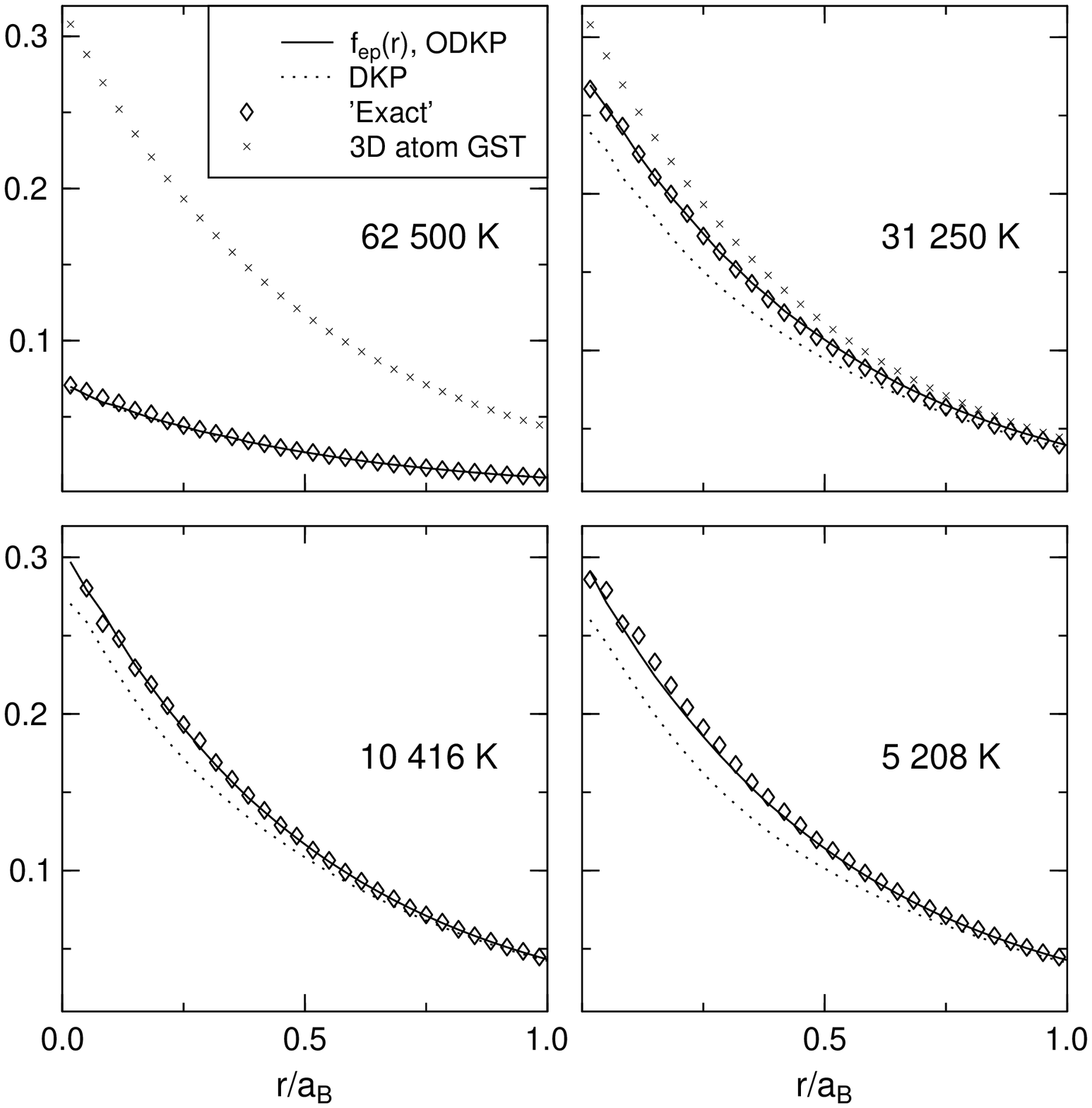}%
\begin{figure}[H]

\vspace{-0.8cm} \caption{Proton-electron pair correlation
functions from PIMC simulations with the ``exact'' pair density
matrix (diamonds), the DKP (dots) and the ODKP (full line).
Temperature values are as indicated in the figure parts:
$T=5\,208,10\,416,\,31\,250$ and $62\,500$ K. For comparison, `3D
GST' -- denotes the pair correlation function corresponding to the
ground state of a hydrogen atom.}

\label{prot_elect_f}
\end{figure}

For temperatures $T=5\,208$ K and $10\,416$ K the hydrogen atom
does not decay into free particles during the duration of a
typical simulation run ($\sim10^{6}$ Monte Carlo steps). In the
figures, the ``exact'' pair correlation function is compared with
the one obtained with the off-diagonal and diagonal Kelbg
potentials, respectively (the number of factorization factors for
the density matrix was chosen to be $M=400$). We found that the
best accuracy is achieved for the off-diagonal Kelbg potential and
$M\gtrsim200$, in this case the ODKP pair correlation function is
very close to the exact one.

At elevated temperatures, $T=31\,250$ K and $62\,500$ K, ionization of
the hydrogen atom occurs, but due to the periodic boundary
conditions, the free particles cannot go to infinity but, when reaching the
boundary, are returned back in the simulation box and have a finite
probability for a formation of a bound state again. Thus, this simulation
captures the region of partial ionization. As the
temperature is increased the ionization probability also increases,
leading to a significant drop in the height of the
proton-electron pair distribution function at the origin compared
to the ground state probability function $\Psi_{0}^{2}(r)$, (see
Fig.~\ref{prot_elect_f}, plots for T=$31\,250$ K and T=$62\,500$ K).

\vspace{-1cm} \hspace{-0.5cm}
\includegraphics[%
  width=9.0cm]{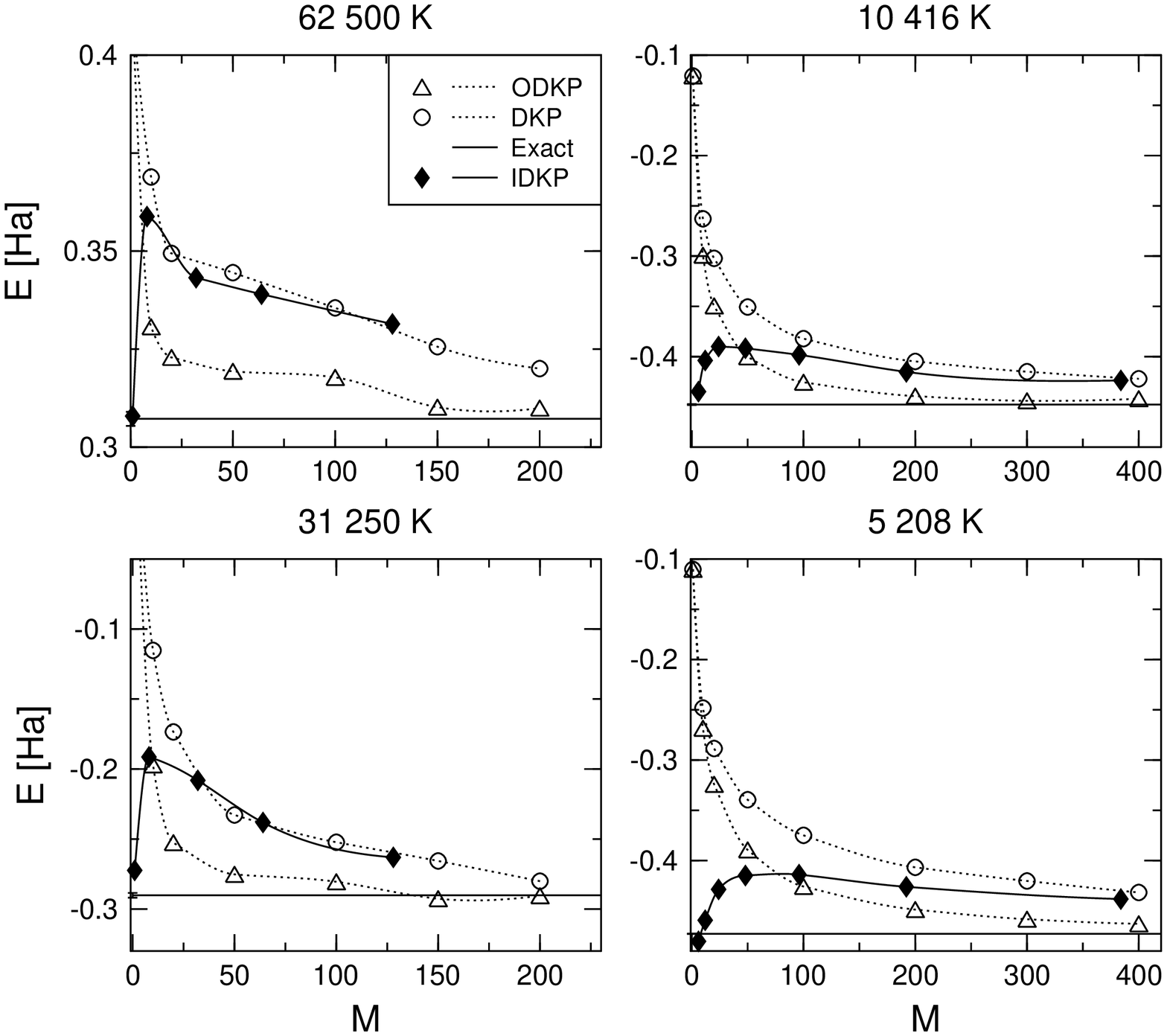}%
\begin{figure}[H]

\vspace{-1.cm} \caption{PIMC results for the internal energy of
the proton-electron pair using ``exact'' pair density matrix, DKP,
ODKP and Improved Kelbg potential vs. different number of
factorization factors $M$.}

\label{energy}
\end{figure}

In Fig.~\ref{energy} we analyze the convergence of the internal
energy in PIMC simulations with varying number of high-temperature
factors $M$. In particular, we compare independent simulations
with the diagonal and off-diagonal Kelbg density matrices,
respectively. The ``exact'' energy value for the considered
temperatures is given by the solid line and is obtained from PIMC
simulations using the ``exact'' pair density matrix, cf.
Sec.~\ref{matsq}). The internal energy was obtained using the
thermodynamic estimator, $\langle E \rangle = -
\frac{\partial}{\partial \beta} \ln Z$, where $Z$ is the partition
function. Comparing the \emph{diagonal} and \emph{off-diagonal}
cases one can note that the ODKP density matrix shows much better
and faster convergence to the exact energy value. A simple
estimate shows that the relative error of the total energy, in the
diagonal approximation, depends on factorization number $M$ as
$\delta E/E\approx\gamma\tau^{2},\;\tau=\beta/M$. In contrast,
using the off-diagonal potential, the error converges much faster,
$\delta E/E\approx\gamma\tau^{3}$.

\begin{figure}[H]
\includegraphics[
  width=8cm, height=9cm]{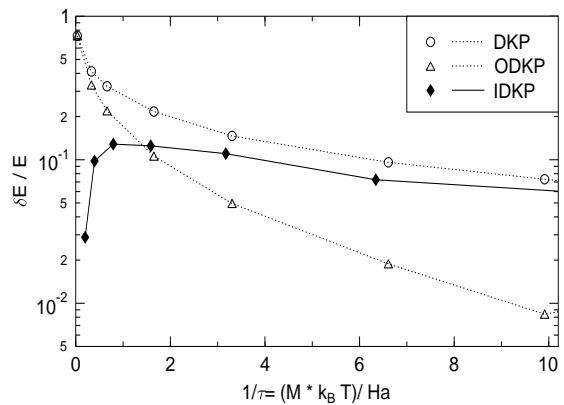}
\vspace{-0.cm} \caption{Relative error in the internal energy of
the proton-electron pair from PIMC simulations with diagonal,
off-diagonal Kelbg potential vs. temperature argument in the
two-particle density matrix $1/\tau$.}

\label{error}
\end{figure}

This fact is illustrated in Fig.~\ref{error} where the logarithm
of the relative error, $\log(\delta E/E)$, is shown as a function
of the inverse of the temperature used in the high-temperature
factors, $1/\tau$. In this figure we compare the behavior of the
error for the same set of temperatures as in Fig.~\ref{energy}. In
Fig.~\ref{energy} we also add simulation results using
\emph{improved diagonal} Kelbg potential (solid line). Its
accuracy is better then that of the ODKP at low temperatures
(small values of $M$) but at high temperatures both are
comparable.

The main conclusion that can be drawn from the presented PIMC
results is that, at equal number of factorization factors $M$,
simulations with the off-diagonal Kelbg potential are
significantly more accurate in reproducing the ``exact''
thermodynamic results of a hydrogen atom. Besides, the full
off-diagonal density matrix contains valuable information about
the spatial electron distribution around the proton, which is lost
in the end-point approximation. Further, we expect that the best
results will be obtained using \emph{an improved off-diagonal}
Kelbg potential, which has the correct zero-point value and
contains the complete angular dependence of the pair density
matrix which, however, is beyond the scope of the present paper.

\vspace{-3.5cm} \hspace{-0.5cm}

\section{Molecular dynamics simulations}
\label{md_results}

In this section, we apply the improved Kelbg potentials in
classical molecular dynamics simulations of dense hydrogen.
Classical MD simulations of dense plasmas have been performed by
many authors before where the divergency at zero distance leading,
in particular, to the classical collapse of an electron into a
proton, is usually avoided by some cutoff or ``regularization'' of
the Coulomb potential at small distances. Using in this paper,
effective quantum pair potentials obtained from the exact
solutions of the Bloch equations, we expect to be on well founded
grounds regarding the correct pair interactions at short
distances. This should not only prevent any collapse but also
correctly reproduce the formation of hydrogen atoms and thus allow
us to obtain essentially improved MD simulation results. On the
other hand, this is not a trivial question, since these potentials
are derived from pure equilibrium considerations, and there is no
ad-hoc proof that they will necessarily be accurate for the
description of dynamical behavior as well, in particular under
strong non-equilibrium conditions. We, therefore, concentrate in
the present analysis on correlated partially ionized hydrogen in
{\em thermodynamic equilibrium}. The results obtained below
confirm that, indeed  (at least in equilibrium), the quantum pair
potentials are well suited for use in the interparticle force
terms in classical MD.

Classical MD simulations incorporate all interparticle collision
processes and are thus not restricted with respect to the coupling
parameter $\Gamma$ in a classical system. With the use of
effective quantum pair potentials, we expect, in addition, to
capture dominant features of the quantum nature of microparticles,
such as quantum diffraction and spin effects. Thus, these
simulations could be called ``semiclassical'' MD. Having access
not only to improved electron-ion potentials but also to
spin-dependent electron-electron potentials, allows us to consider
also spin effects by simulating electrons with different spin
projections as two independent particle species.  No spin-flip
processes, as they would be expected e.g. in strong magnetic
fields are considered \cite{spinflip}, but our model should be capable to describe
related phenomena as well. In this paper we focus on static
properties, such as internal energy, and radial distribution
functions. Investigation of dynamical properties an of spin
density fluctuation is the aim of a forthcoming paper.

We consider a dense, degenerate hydrogen plasma at two densities
corresponding to the Brueckner parameter $r_s={\bar r}/a_B=4$ and $r_s=6$
and temperatures $T= 31\, 250, 50\, 000, 62\, 500,
125\, 000$ and $166\, 670$~K. These parameter values correspond,
respectively, to
$\Gamma$= 2.53, 1.58, 1.26, 0.63 and 0.47, for $r_{s}$= 4, and 1.68,
1.05, 0.84, 0.42, 0.32, for $r_{s}$= 6.

The simulation box of our system, with the length
$L$=$[n/(N_{p}+N_{e}^{\uparrow}+N_{e}^{\downarrow})]^{1/3}$,
contains $N_{p}$=~200 protons, $N_{e}^{\uparrow}$= 100
electrons with spin up and an equal number of electrons,
$N_{e}^{\downarrow}$= 100, with spin down. We keep the condition of
the electro neutrality by taking $N_{p}$= $N_{e}^{\downarrow}$+
$N_{e}^{\uparrow}$. Details of the used numerical algorithm can be
found in Ref.~\cite{vova01}.

Since MD, in contrast to PIMC, involves only diagonal interaction
potentials, we choose the following expressions:
for the interaction between electrons and protons, protons and protons
and electrons with opposite spin, we use the improved Kelbg potential,
Eq.~(\ref{impr_kelbg}), with the fit parameters given by
Eqs.~(\ref{ep_fit_gamma}) and~(\ref{ee_up_down_fit_gamma}),
respectively.
The interaction between
electrons with the same spin projection is described by the diagonal
antisymmetric potential, Eq.~(\ref{u_e_d}).
Further, to properly account for the long range-character of the potentials,
we used the standard Ewald procedure as in Ref.~\cite{vova01} whereas, in
contrast to the rather involved expressions given there for a one-component
plasma, here we could restrict the potential energy sum only by the proper
sum in {\em real space} (we do not reproduce these lengthy expressions here,
but mention that the value of the parameter $\alpha$ defined in Ref.~\cite{vova01}
was chosen to be $\alpha$ = $5.6/L$)
and taking 5 vectors in every direction in {\em the reciprocal space}. This
gives some computational-cost advantage in computation of the
forces compared to Ref.~\cite{vova01}.

\begin{figure}
\includegraphics[width=6.2cm,angle=-90]
{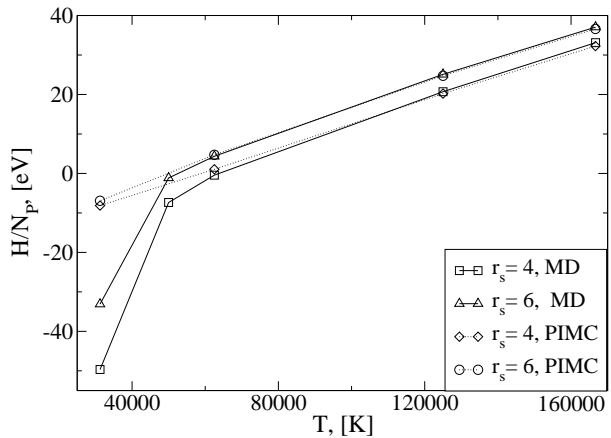}\vspace{-0.cm}
\caption{Semiclassical MD results (full lines) for the internal
energy per hydrogen atom  at densities $r_{s}$= 4, 6 versus
temperature. The results of restricted PIMC simulations by
Militzer~\cite{MilitzerPhD} are shown for comparison (dashed
lines). Symbols indicate the five temperatures for which MD
simulations have been performed: $T= 31\, 250$, $50\, 000$, $62\,
500$, $125\, 000$ and $166\, 670$~K (solid lines). The pair
approximation breaks down around $50\,000K$, at the molecule
binding energy. \label{cap:Internal-energy}}
\end{figure}

In Fig.~\ref{cap:Internal-energy}, we plot the internal energy per
atom as a function of temperature for two densities and compare it
to the path integral Monte Carlo results of Militzer
\cite{MilitzerPhD}. One can note, that for high temperatures the
energies of MD and PIMC simulations coincide very well and lie
within the limits of the statistical errors. This is an important
test for the simulation, and this agreement was expected due to
the asymptotic character of the Kelbg potential as a rigorous weak
coupling result. Moreover, we observe practically full agreement
between MD and PIMC results to temperature as low as approximately
$50\,000$K, corresponding to a coupling parameter $\Gamma=3$. This
is a remarkable extension of ``semiclassical'' molecular dynamics
into the regime of moderate coupling and moderate degeneracy.

Naturally, below a critical temperature of about $50\,000$~K
deviations from PIMC results start to grow rapidly -- the MD
results for the energy are becoming very low. It is very
interesting to analyze the reasons for these deviations, as this
may suggest directions for further improvements. It turns out that
the reason for these deviations is not a failure of the used
quantum pair potentials. Thus the only source for the deviations
in the full simulation can be many-particle effects beyond the
two-particle level.

To verify this hypothesis we performed a careful inspection of the
microscopic particle configurations in the simulation box. At high temperatures,
the particle trajectories are those of a fully ionized classical plasma.
At temperatures below on Ry, we observe an increasing number of electrons
undergoing strong deflections on protons and eventually performing quasibound
trajectories. Most of these electrons remain ``bound'' only for a few classical orbits
and then leave the proton again. Averaged over a long time our simulations are
able to reveal the degree of ionization of the plasma. At the same time we
observe occasional events of three and more particles being at short
distances for the duration of one or more orbits, reflecting the appearance
of hydrogen molecules $H_2$, molecular ions $H_2^+$ etc.

If the temperature is lowered below approximately $T=50\,000$~K,
we observe a strong increase of molecule formation and even an
aggregation of many molecules into clusters with an interparticle
distance close to one $a_B$. This turns out to be the reason for
the observed very low energy because the attractive Coulomb
interaction contributions are becoming dominant in the total
energy. Of course, this behavior is not surprising: while all pair
interaction processes are modelled correctly even at low
temperature (which is assured by the fit parameters in the
improved Kelbg potentials), as soon as three or more particles are
being closely together, three-particle and higher order
correlations are becoming increasingly strong (they, in
particular, account for the formation of the larger bound state
complexes described above). However, it was just the approximation
used in the derivation of the quantum potentials that
three-particle and higher correlations can be neglected which was
the basis for the use of pair potentials in modelling the whole
$N$-particle system. While molecular dynamics, of course, includes
any level of correlations, the use of the present potentials means
that {\em quantum corrections to three-body (and higher order)
interactions} are not adequately captured. Therefore, it is no
surprise that this approximation breaks down at sufficiently low
temperature, and this break down occurs around the temperature
corresponding to the binding energy of hydrogen molecules. From
this we can conclude that molecule formation sets the limit of the
applicability of the present semiclassical MD simulations.

%\vspace{-0.2cm}\hspace{-0.3cm}

Let us now turn to a more detailed analysis of the spatial
configuration of the particles in the MD simulations.

%\vspace{-0.2cm}
\begin{figure}
\includegraphics[angle=-90,width=10cm]{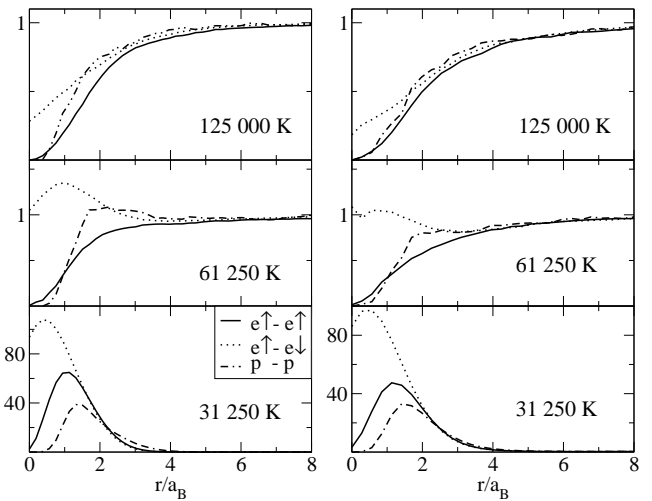}
\vspace{0cm} \caption{Electron-electron and proton-proton radial pair
distribution functions for a correlated
hydrogen plasma with $r_{s}$= 4 (left row) and $r_{s}$= 6 (right row) for
$T = 125\, 000, 61\, 250$ and $31\,250$~K (from top to
bottom).\label{cap:Radial-distribution-functions}}
\end{figure}

In  Fig.~\ref{cap:Radial-distribution-functions} the radial pair
distributions between all particle species with the same charge
are plotted at two densities. Consider first the case of $T=125\,
000$~K (upper panel). For both densities, all functions look
qualitatively the same, showing a depletion at zero distance due
to Coulomb repulsion. Besides, there are differences which arise
from the spin properties. Electrons with same spin show a slightly
broader ``Coulomb hole'' around $r=0$ than the protons, because
the Pauli principle yields an additional repulsion of the
electrons (this effect is much weaker for two protons due to their
much smaller de Broigle wavelength). This trend is reversed at
lower temperature, see middle panel, which is due to the formation
of hydrogen atoms, see also Fig.~\ref{cap:g_r2} below. In this
case, electrons (their trajectories) are ``spread out'' around the
protons giving rise to an increased probability of close
encounters of two electrons in different atoms compared to two
protons.

Now, let us compare electrons with parallel vs. electrons with antiparallel
spins. In all cases, we observe a significantly increased probability
to find two electrons with opposite spin and small distances below one
Bohr radius which is due to the missing Pauli repulsion in this case.
This trend increases with lowering of temperature due to increasing
quantum effects and thus convincingly confirms that spin effects are
correctly reproduced in our MD simulations.

%\vspace{-0.3cm}

\begin{figure}
\includegraphics[
  width=6.5cm,
  angle=-90]{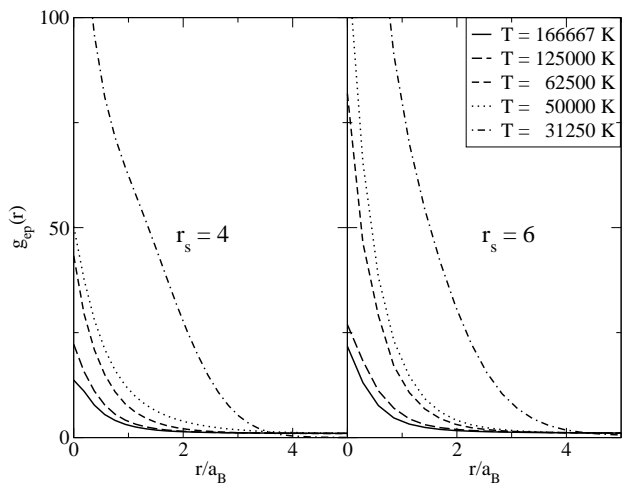}\vspace{-0.cm}
\caption{Electron-proton radial pair distribution functions at
$r_{s}$= 4 (left figure) and $r_{s}$= 6 (right figure) and five
temperatures: $T= 166\, 667, 125\, 000, 62\, 500$ and $50\, 000$~K.
\label{cap:g_ep}}
\end{figure}

Before analyzing the lowest temperature in
Fig.~\ref{cap:Radial-distribution-functions} let us consider the
electron-proton pair distributions which are shown in
Fig.~\ref{cap:g_ep}. With lowering of temperature, we observe a
strong increase of the probability to find an electron close to a
proton. In contrast to the classical case of a collapse (see
above), here this probability is finite. Multiplying these
functions by $r^2$ gives essentially the radial probability which
is plotted in Fig.~\ref{cap:g_r2}. Here, lowering of temperature
leads towards formation of shoulder around $1.5a_B$ which is due
to the formation of hydrogen atoms. This conclusion is confirmed
by inspection of the corresponding quasibound electron
trajectories as discussed above. At this temperature, the observed
most probable electron distance is not $1a_B$ as in the hydrogen
ground state which is due to the considerable kinetic energy of
the particles leading to a larger average radius of the classical
quasiclosed orbits We expect that at lower temperature, the most
probable radius would tend towards $1a_B$, but this temperature
range is not realistically modelled due to molecule and cluster
formation.

\vspace{-0.cm}

\includegraphics[%
  width=6.7cm,
  angle=-90]{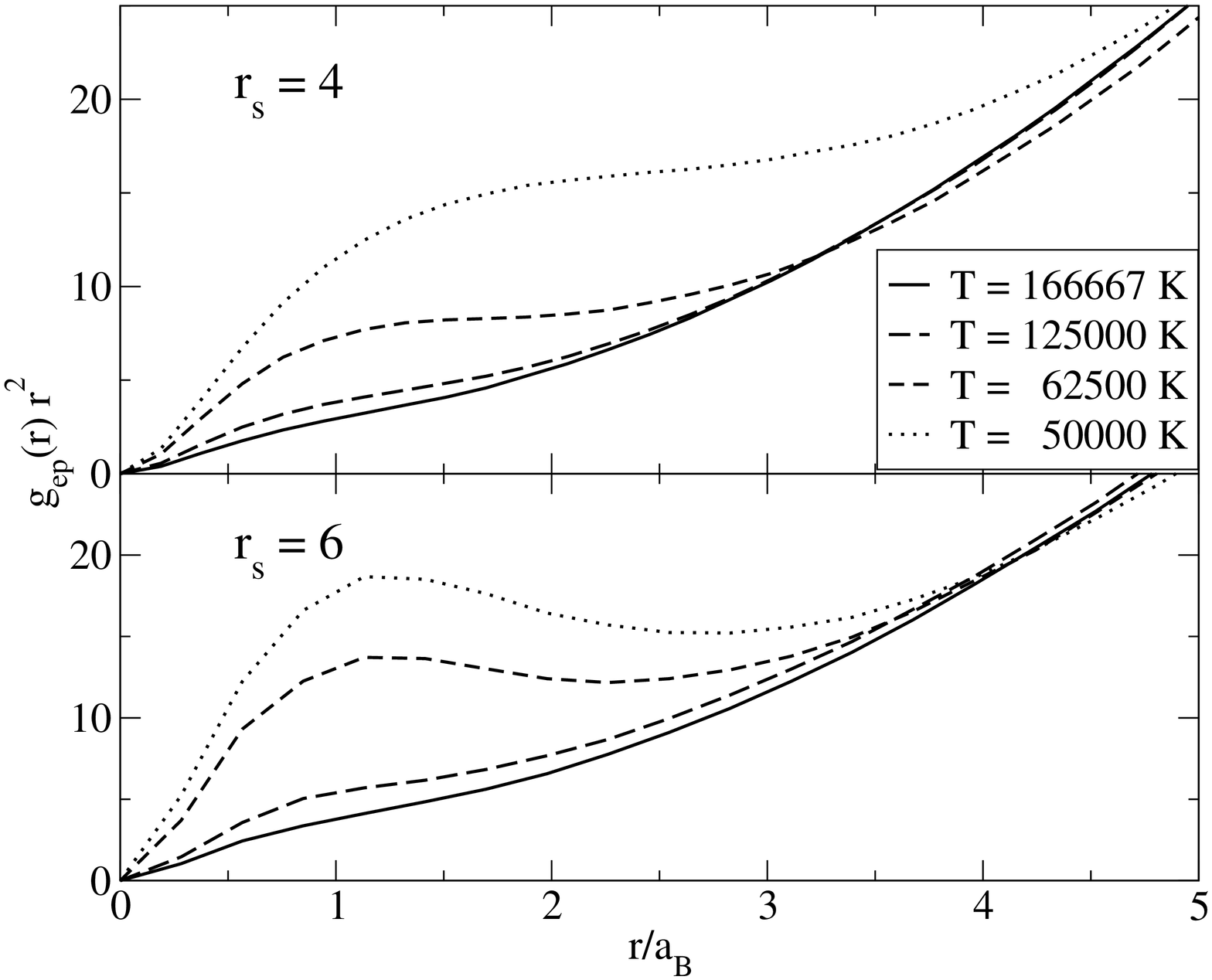}\vspace{-0.5cm}%
\begin{figure}[H]

\caption{Electron-proton radial pair distribution functions
multiplied by $r^{2}$. Same data as in Fig.~\ref{cap:g_ep}.
\label{cap:g_r2}}
\end{figure}
While the description of correlated complexes of more than two particles
is certainly beyond the present pair approximation model, nevertheless,
several features of a partially ionized and partially dissociated
hydrogen plasma are reproduced correctly. At $62\, 500$~K and $r_s=6$
(right center part of Fig.~\ref{cap:Radial-distribution-functions})
the simulations show a first weak signature of molecule formation --
see the maximum of the p-p pair distribution function around $r=2a_B$
and the maximum of the pair distribution function of electrons
with antiparallel spins around $r=1.5a_B$. Further lowering of the
temperature by a factor of two
(lower panel of Fig.~\ref{cap:Radial-distribution-functions})
confirms this trend: the p-p functions exhibit a clear peak very close
to $r=1.4a_B$ -- the theoretical p-p separation in $H_2$-molecules.
At the same time, also the e-e functions have a clear peak around
$r=0.5a_B$, in the case of opposite spins, and $r=1.2a_B$, for
parallel spin projections. The first case comes rather close to the
true quantum mechanical $H-H$ bound state (singlet) with the electron
wave function predominantly concentrated between the two protons.
On the other hand, this electron peak should also extend to the right
of the p-p peak, and no such pronounced peak would be expected for
electrons with the same spin.

Nevertheless, we may conclude that even formation and spatial
dimension of hydrogen molecules appear to be captured surprisingly
well in these simulations. The main difficulty appears to arise
not on the level of four-particle correlations but on the level of
six particle correlations: in the simulations nothing prevents two
``bound'' atoms from binding to a third and more atoms. The
overall attractive Coulomb interaction makes it, below
$50\,000$~K, energetically favorable to form large clusters
consisting of more than two atoms, explaining the strong decrease
of the internal energy at the $T=31\, 250$~K, cf.
Fig.~\ref{cap:Internal-energy}. In reality, complexes of two
molecules do exist, but they have a very low binding energy which
is due to a subtle compensation effects arising from repulsive
exchange interaction between the electrons which go far beyond the
level of pair interactions \cite{vova_phd}.

\section{Use of the quantum pair potentials in density functional theory}\label{dft}

The effective quantum potentials have been introduced to represent the
equilibrium two particle density matrix and subsequently generalized to
incorporate many-body Coulomb coupling effects. There are other many body
coupling effects due to degeneracy or exchange correlations. For some
applications, it may be useful to incorporate these directly in
the effective pair potentials to extend their validity to still lower
temperatures, as was demonstrated on the example of classical MD above.

 In this section, we describe the usefulness of the effective
quantum potentials for a completely different theoretical approach
-- density functional theory (DFT). In doing so, the role of
effective quantum potentials with degeneracy effects is
illustrated as well.

DFT is a formal structure in which non-perturbative approximations
can \ be introduced to describe strong coupling effects
\cite{mermin}. Although there are both classical and quantum
versions of DFT, the classical form does not apply to a system of
electrons and positive ions due to the Coulomb divergence. One
possibility is to postulate a classical statistical mechanics
using the effective quantum potentials described above which
allows to remove the singularity. Alternatively, the proper
quantum formulation can be used from the outset and the effective
quantum potentials ``derived'' as a tool in the process of
computing properties of interest \cite{zogaib}. This second
approach will be used here.

In essence, DFT is a variational means to derive an equation
for the charge density induced by an external potential. If that potential
is taken to be the same as the potential of one of the charges in the system, the resulting
density is in fact formally identical to the equilibrium pair correlation function, or diagonal
element of the two particle density matrix.
The density obeys a known nonlinear integral equation -- a generalization of the
Boltzmann-Poisson equation. However, in practice, the direct solution
this equation is seldom attempted. Instead an equivalent set
of self-consistent one particle Schr\"odinger equations, the Kohn-Sham
equations \cite{ks}, are solved first to construct the charge density.
Yet it might be very useful to recall the existence of an alternative direct approach
which becomes practical if an appropriate quantum
pair potential is introduced. This is illustrated in more detail as follows.

Consider a quantum system in the presence of external sources that can be
described by an additive potential
\begin{equation}
\widehat{V}=\sum_{\alpha }\sum_{i=1}^{N_{\alpha }}\widehat{V}_{\alpha }({\bf %
q}_{i\alpha }).  \label{1.1}
\end{equation}%
Here $\alpha $ denotes a species and ${\bf q}_{i\alpha }$ is the
position operator of particle $i$ of species $\alpha $. The caret
on the potentials is used to distinguish the quantum operator from
its corresponding function. In general, each species may have a
different form for the coupling to the external sources. The
potential also can be written in terms of the density operators
for each species
\begin{equation}
\widehat{V}=\int d{\bf r}\sum_{\alpha }V_{\alpha }({\bf r})\widehat{n}
_{\alpha }({\bf r}),\hspace{0.25in}\widehat{n}_{\alpha }({\bf r}
)=\sum_{i=1}^{N_{\alpha }}\delta ({\bf r-q}_{i\alpha }).  \label{1.2}
\end{equation}%
The details of the remainder of the Hamiltonian are not important at this
point. For this many-body system with external sources the theorems of
density functional theory apply in the following form. First, a functional
of the average densities, $\widehat{n}_{\alpha }({\bf r})$, averaged over an
equilibrium grand canonical ensemble is constructed (the generalization to
other equilibrium ensembles has been carried out). This is done in two steps.
First, the equilibrium grand potential for the system is considered formally
\begin{equation}
\beta \Omega _{e}=-\ln \sum_{\left\{ n_{\alpha }\right\} }
{\rm Tr}\, e^{-\beta \left(
H-\sum_{\alpha }\mu _{\alpha }n_{\alpha }\right) }.  \label{1.3}
\end{equation}%
The density for the various species is obtained (formally) by functional
differentiation with \ respect to the potentials
\begin{equation}
\Omega _{e}=\Omega _{e}\left( \left\{ \mu _{\alpha }-V_{\alpha }\right\}
\right) , \; n_{e\alpha }\left( r\right) =-\frac{\delta \Omega _{e}
}{\delta \left[ \mu _{\alpha }-V_{\alpha }(r)\right] }.  \label{1.4}
\end{equation}%
The density equation is inverted (formally) to get the external potentials
as functionals of the average densities
\begin{equation}
V_{\alpha }={\cal V}_{\alpha }({\bf r}\mid \left\{ n_{e\sigma }\right\} ),
\label{1.5}
\end{equation}%
and a Legendre transformation is performed to construct the free energy as a
functional of the densities rather than the chemical potentials
\begin{eqnarray}
F(\left\{ n_{e\alpha }\right\} )
&=&\Omega _{e}\left( \left\{ \mu _{\alpha }-{\cal V}_{\alpha }\right\} \right)
\\\nonumber
&+&
\sum_{\alpha }\int d{\bf r}\left[ \mu
_{\alpha }-{\cal V}_{\alpha }\left( {\bf r\mid }\left\{ n_{e\sigma }\right\}
\right) \right] n_{e\alpha }\left( {\bf r}\right).   \label{1.6}
\end{eqnarray}
The crucial second step is to extend this functional to {\em arbitrary
density fields}
\begin{equation}
F(\left\{ n_{e\alpha }\right\} )\rightarrow F(\left\{ n\right\} ).
\label{1.7}
\end{equation}%
The main task of density functional theory is now to
construct the density functional
\begin{equation}
\Omega _{V}\left( \left\{ n\right\} \right) \equiv F(\left\{ n\right\}
)-\int d{\bf r}\left( \mu _{\alpha }-V_{\alpha }\left( {\bf r}\right)
\right) n_{\alpha }\left( {\bf r}\right),
\end{equation}%
where, in this definition, $ V_{\alpha }\left( {\bf r}\right) $
is {\em not} considered to be a functional of the density. The
main theorem of density functional theory is then that this functional has
an extremum at the equilibrium density
\begin{eqnarray}
\frac{\delta \Omega _{V}\left( \left\{ n\right\} \right) }{\delta n}=0
&=& \frac{\delta F\left( \left\{ n\right\} \right) }{\delta n}-
\left[ \mu _{\alpha}-V_{\alpha }\left( {\bf r}\right) \right],
\nonumber\\
\Rightarrow n &=& n_{e\alpha }.
\label{1.9}
\end{eqnarray}
Furthermore the value of the functional at the equilibrium density is
clearly the equilibrium grand potential%
\[
\Omega _{V}\left( \left\{ n\right\} \right) =\Omega \left( \left\{ \mu
_{\alpha }-{\cal V}_{\alpha }\right\} \right) .
\]%
In practice, an approximate free energy functional $F\left( \left\{
n\right\} \right) $ is written and Eq. (\ref{1.9}) is solved to obtain the
equilibrium density. This density is then used to evaluate the equilibrium
grand potential and determine all equilibrium thermodynamic properties.
Structural properties can be obtained as well by choosing the external
potential at the end to be the same as that for interaction among the system
particles. In other words, the source is chosen to be a particle of the same
type as those comprising the many-body system. The densities $n_{e\alpha }$
become equilibrium pair correlation functions.

How should the functional $F(\left\{ n\right\} )$ be constructed? There is
clearly a part associated with an ideal gas, and an energy due to the direct
Coulomb interactions. These can be identified explicitly. In addition there
are the more difficult parts due to exchange and correlations. Consequently,
it has become standard practice to write the free energy as
\begin{eqnarray}
F[n] &=& F^{(0)}(\left\{ n\right\} ) +
\\\nonumber
&& \frac{1}{2}\sum_{\alpha ,\sigma }
\int d{\bf r}d{\bf r}^{\prime }V_{\alpha \sigma }({\bf r}-{\bf r}^{\prime })
n_{\alpha}({\bf r}) n_{\sigma}({\bf r}^{\prime})
+F_{xc}(\left\{ n\right\} ),  \label{1.10}
\end{eqnarray}
where $F^{(0)}(\left\{ n\right\} )$ is the free energy for the
non-interacting system, the second term is the contribution from
the direct Coulomb interaction, and $F_{xc}(\left\{ n\right\} )$
denotes the remaining contributions due to interactions from
exchange and correlations. Then the extremum condition (\ref{1.9})
becomes \cite{zogaib}
\begin{eqnarray}
&&{\cal V}_{\alpha }^{(0)}\left( {\bf r\mid }\left\{ n_{\sigma }\right\}
\right) = V_{\alpha }\left( {\bf r}\right) +
\label{1.11}\\\nonumber
&& \quad\sum_{\sigma }\int d{\bf r}d{\bf r}^{\prime }V_{\alpha \sigma }
({\bf r}-{\bf r}^{\prime })n_{\sigma }\left(
{\bf r}^{\prime }\right) +\frac{\delta F_{xc}\left( \left\{ n\right\}
\right) }{\delta n_{\alpha }\left( {\bf r}\right) },
\end{eqnarray}
with ${\cal V}_{\alpha }^{(0)}({\bf r}\mid n)$ denoting the functional
(\ref {1.5}) for the ideal gas.
Determination of this functional is the central
issue of the discussion here, and we will show that it is closely related to
the Kelbg potential analyzed in the bulk of this paper.

The definition of the functional ${\cal V}_{\alpha }^{(0)}({\bf r}\mid n)$
is straightforward from the
representation of the density for an {\em ideal Fermi gas} in the external potentials
\begin{equation}
n_{\alpha }(r) =\left\langle r\right| \left( e^{\beta \left(
\frac{p^{2}}{2m_{\alpha }}+\widehat{V}_{\alpha }-\mu _{\alpha }\right)
}+1\right) ^{-1}\left| r\right\rangle .  \label{1.12}
\end{equation}
This is a single particle problem. The right side is clearly a functional of
$V_{\alpha }$ through the dependence of the eigenvalues of $\frac{p^{2}}{%
2m_{\alpha }}+\widehat{V}_{\alpha }$ on the form of the external potential.
Interestingly, even at the level of the ideal gas determination of this
functional is non-trivial. In the {\em non-degenerate limit}
this equation for the density becomes
\begin{equation}
n_{\alpha }\left( r\right) \rightarrow \left\langle r\right| e^{-\beta
\left( \frac{p^{2}}{2m_{\alpha }}+\widehat{V}_{\alpha }-\mu _{\alpha
}\right) }\left| r\right\rangle .  \label{1.13}
\end{equation}%
If the external potential is chosen to be a Coulomb source, then (\ref{1.13})
becomes equivalent to the diagonal elements of the two particle density
matrix in relative coordinates which has exactly the form of the
pair distribution function used to define the effective quantum pair
potential, cf. Eq.~(\ref{exact_diag_u}).

Once the exchange and correlation free energy functional is specified
(guessed), (\ref{1.11}) provides a set of closed {\em classical} integral
equations for the equilibrium densities. As will \ be seen below, a leading
approximation is the usual Boltzmann-Poisson representation in terms of
semi-classical potentials. The primary technical difficulty in this
prescription is the determination of ${\cal V}_{\alpha }^{(0)}({\bf r}\mid n)
$. Kohn and Sham noted that (\ref{1.11}) defines an effective single
particle potential and therefore is formally equivalent to the ideal gas in
this effective potential. Therefore, the solution can be constructed by
solving the one particle Schr\"odinger equation whose potential is the right
side of (\ref{1.11}), and calculating the densities from the associated form
(\ref{1.12}) self-consistently
\begin{eqnarray}
n_{\alpha }\left( r\right) &=&
\left\langle r\right| \left( e^{\beta \left(
\frac{p^{2}}{2m_{\alpha }}+\widehat{V}_{\alpha }-\mu _{\alpha }\right)
}+1\right) ^{-1}\left| r\right\rangle
\nonumber\\
&=&\sum_{i}\left( e^{\beta \left(
\epsilon _{i\alpha }-\mu _{\alpha }\right) }+1\right) ^{-1}\left| \psi
_{i}\left( r\right) \right| ^{2}.  \label{1.14}
\end{eqnarray}
This avoids the difficult problem of finding the functional ${\cal V}%
_{\alpha }^{(0)}({\bf r}\mid n)$ but at the cost of having to solve a set of
self-consistent Schr\"odinger equations.

Consider instead an approximate evaluation of the potential
${\cal V}_{\alpha }^{(0)}({\bf r}\mid n)$ in terms of an effective
quantum potential $U_{\alpha }(r)$
defined by
\begin{eqnarray}
n_{\alpha}(r)
& \equiv &
\int \frac{d{\bf p}}{(2\pi \hbar)^{3}}
\left( e^{\beta \left( \frac{p^{2}}{2m_{\alpha }}
+U_{\alpha }(r)-\mu _{\alpha }\right) }+1\right)^{-1}
\nonumber\\
&=& \left\langle r\right|
\left( e^{\beta \left( \frac{\widehat{p}^{2}}{2m_{\alpha }}+\widehat{{\cal V}%
}_{\alpha }^{(0)}-\mu _{\alpha }\right) }+1\right) ^{-1}\left|r\right\rangle.
\label{1.15}
\end{eqnarray}
The first equality is similar to a finite temperature Thomas-Fermi
representation, with a local chemical potential given by $\mu _{\alpha
}\left( r\right) =\mu _{\alpha }-U_{\alpha }(r)$. An important difference
discussed below is that $U_{\alpha }(r)\neq V_{\alpha }\left( r\right) $.
The functional relationship of $n_{\alpha }\left( r\right) $ to $\mu
_{\alpha }\left( r\right) $ and hence to $U_{\alpha }(r)$ is that for an
ideal gas and is well-known. The second equality of (\ref{1.15}) defines the
semi-classical potential $U_{\alpha }(r\mid {\cal V}_{\alpha }^{(0)})$ as a
functional of ${\cal V}_{\alpha }^{(0)}$.  This relationship of $U_{\alpha
}(r\mid {\cal V}_{\alpha }^{(0)})$ to ${\cal V}_{\alpha }^{(0)}$ is more
difficult to unfold. However, it is straightforward to discover it for weak
coupling of the system to the perturbing potential. The analysis is similar
to the derivation of the Kelbg potential and will not be repeated here.
Formally make the replacement $\widehat{{\cal V}}_{\alpha }^{(0)}\rightarrow
\lambda \widehat{{\cal V}}_{\alpha }^{(0)}$ in (\ref{1.15}) with the
corresponding dependence on $\lambda $ inherited by $U_{\alpha }(r)$. Then
perform the expansion of $U_{\alpha }(r)$ to first order in $\lambda $,
setting $\lambda =1$ at the end, to get \cite{zogaib}
\begin{equation}
U_{\alpha }(r)\rightarrow \int d{\bf r}^{\prime }\pi _{\alpha }\left( {\bf r}%
-{\bf r}^{\prime }\right) {\cal V}_{\alpha }^{(0)}({\bf r}^{\prime })
\label{1.16}
\end{equation}%
where $\pi _{\alpha }\left( {\bf r},{\bf r}^{\prime }\right) $ is the well-known
static linear polarization function in random phase approximation,
\begin{eqnarray}
\pi _{\alpha }\left( {\bf r}\right) &=& ( 2\pi ) ^{-3}
\int d{\bf r} \: e^{i{\bf k\cdot r}}\,\widetilde{\pi }_{\alpha }\left( k\right),
\\
\widetilde{\pi }_{\alpha }\left( k\right)
&=& \frac{\partial \mu _{\alpha }}{\partial n_{\alpha }}
\int \frac{d{\bf p}}{(2\pi\hbar)^{3}}
\frac{F_{\alpha }({\bf p}-\hbar{\bf k}) - F_{\alpha }({\bf p}) }
{p^{2} - ( {\bf p}-\hbar {\bf k}) ^{2}},
\label{1.17}
\end{eqnarray}
containing the Fermi distribution
\begin{equation}
F_{\alpha}\left( p\right) =\left( e^{\beta \left( \frac{p^{2}}{2m_{\alpha }}-\mu
_{\alpha }\right) }+1\right) ^{-1}.  \label{1.18}
\end{equation}%
In this approximation, the functional relationship between the density and
the potential is now known%
\begin{equation}
n_{i}\left( r\right) \equiv
\int \frac{d{\bf p}}{(2\pi \hbar)^{3}}
\left( e^{\beta \left( \frac{p^{2}}{2m_{i}}+\int d%
{\bf r}^{\prime }\pi \left( {\bf r}-{\bf r}^{\prime }\right) {\cal V}%
_{i}^{(0)}\left( r^{\prime }\right) -\mu _{i}\right) }+1\right) ^{-1}
\label{1.19}
\end{equation}
Now it is straightforward to improve this results by substitution of (\ref{1.11})
into the right side of (\ref{1.19}) which gives a
generalization of the Thomas-Fermi approximation to include strong coupling
effects. However, even if $F_{xc}\left( \left\{ n\right\} \right) $ is
neglected the result is the Thomas-Fermi approximation in terms of the
potential
\begin{equation}
\overline{V}_{\alpha }({\bf r})=\int d{\bf r}^{\prime }\pi _{\alpha }\left(
{\bf r}-{\bf r}^{\prime }\right) V_{\alpha }({\bf r}^{\prime })
\label{1.19a}
\end{equation}
rather than the bare potential $V_{\alpha }({\bf r})$, which has short
ranged divergences for opposite charge interactions. The result here in
terms of the {\em nonlocal effective quantum  potential} appears to be a new one
that cures some of the well-known problems of the ``local approximation''
Thomas-Fermi theory. As indicated below,  $\overline{V}_{\alpha }({\bf r})$
becomes just the Kelbg potential in the non-degenerate limit.
The result (\ref{1.19}) with (\ref{1.11}) is a non-linear integral equation
for the density, including both {\em strong coupling and degeneracy effects}.
There is no longer any need to solve the Kohn-Sham equations and the problem
is one of purely classical analysis.

It is instructive to consider the
non-degenerate limit. In that case the polarization function is evaluated
using  $F_{0}\left( p\right) \rightarrow
e^{-\beta \left( \frac{p^{2}}{2m_{\alpha }}-\mu _{\alpha }\right) }$.
Furthermore, Eq.~(\ref{1.19}) simplifies to
\begin{eqnarray}
n_{\alpha }\left( r\right) &=&
n_{\alpha }e^{-\beta U_{\alpha }(r)},
\label{1.20a}\\
{\cal V}_{\alpha }^{(0)}\left( r^{\prime }\right) &=&
\int d {\bf r}^{\prime}\,\pi _{\alpha }^{-1}\left( {\bf r}-{\bf r}^{\prime }\right)
U_{\alpha }\left( r^{\prime }\right) .  \label{1.20}
\end{eqnarray}
Use of these in the DFT equation (\ref{1.11}) gives the closed equation for
the densities
\begin{eqnarray}
\ln \frac{n_{\alpha }\left( r\right) }{n_{\alpha }} &=&
-\beta \overline{V}
_{\alpha }\left( {\bf r}\right) - \beta \sum_{\sigma }
\int d{\bf r}d{\bf r}^{\prime }\,\overline{V}_{\alpha \sigma }({\bf r}-{\bf r}^{\prime })
n_{\sigma}\left( {\bf r}^{\prime }\right)
\nonumber\\
&+&\int d{\bf r}^{\prime }\pi _{\alpha}\left( {\bf r}-{\bf r}^{\prime }\right)
\frac{\delta F_{xc}\left( \left\{
n\right\} \right) }{\delta n_{\alpha }\left( {\bf r}^{\prime }\right) }.
\label{1.21}
\end{eqnarray}
The potentials $\overline{V}_{\alpha }\left( {\bf r}\right) $ and
$\overline{V}_{\alpha \sigma }({\bf r}-{\bf r}^{\prime })$ are ``regularized'' by the
polarization function, e.g.,
\begin{equation}
\overline{V}_{\alpha }\left( {\bf r}\right) =\int d{\bf r}^{\prime }\pi
_{\alpha }\left( {\bf r}-{\bf r}^{\prime }\right) V_{\alpha }\left( {\bf r}%
^{\prime }\right) .  \label{1.22}
\end{equation}
It is possible to show \cite{dufty} that in this {\em non-degenerate limit} $\overline{V}
_{\alpha }\left( {\bf r}\right) $ {\em is just the original Kelbg potential},
Eq.~(\ref{kelbg-d}). Therefore, in
the weak coupling limit where $F_{xc}\left( \left\{ n\right\} \right) $ can
be neglected (\ref{1.21}) becomes the usual Boltzmann-Poisson equation with
effective quantum potentials given by the Kelbg potential (\ref{kelbg-d}).

In summary, DFT applications can be performed in a semi-classical form
without solving the Kohn-Sham equations by introducing effective quantum
potentials. This can be done in a weak coupling approximation similar to
that first described by Kelbg and yields a closed analytical result (\ref{kelbg-d}).
Based on the results of the above analysis, it can be expected that this
approach can be extended by incorporating as well effects of
degeneracy by using for the density Eq.~(\ref{1.12}) instead of (\ref{1.20a}).
Furthermore by using {\em improved quantum pair potentials} -- along the lines of the
improved Kelbg potentials discussed in the previous sections -- an accurate
treatment of the pair problem is achieved laying the foundation for
advancing DFT to the regime of strong coupling.

\section{Conclusion}
\label{dis}

In this work we presented an analysis of generalized quantum pair
potentials. Extending the work of Kelbg and others we investigated in
detail {\em effective off-diagonal and diagonal quantum pair potentials} for
a correlated hydrogen plasma including spin effects.
We studied the accuracy of these potentials by an extensive comparison
with the exact solutions of the Bloch equation.
Further, we proceeded to an analysis of {\em improved diagonal quantum pair
potentials} by correcting the value of the Kelbg potential at zero
particle separation. Excellent agreement with the exact solutions of the
two-particle Bloch equations could be achieved with the help of a single
temperature-dependent fit parameter for which an accurate
analytical Pad\'e formula was presented. This lead to significantly
improved diagonal pair potentials compared to the original Kelbg potential.
Moreover, these potentials are explicitly spin-dependent and retain
the advantage of a closed analytical expression.

These
potentials have been applied in path integral Monte Carlo and
``semiclassical'' molecular dynamics simulations of dense hydrogen and
were found to give accurate results over a wide range of parameters.
One important conclusion, of relevance to PIMC simulations, is
that the off-diagonal potential gives essentially more accurate results
(or more rapid convergence) than its
diagonal limit, quantitative estimates have been provided.

Furthermore, we have demonstrated that the spin-dependent improved diagonal
potentials are of high use for ``semiclassical'' molecular dynamics
simulations of partially ionized plasmas. Our analysis revealed that with
these potentials one can successfully simulate dense hydrogen
up to moderate coupling and degeneracy, from the fully ionized to the
partially ionized regime. The present potentials
allow us to correctly model dense plasmas up to temperatures as low as
the molecular binding energy. Further improvements are possible, including
the description of molecular hydrogen, but they require to include three-particle
and four-particle correlations and exchange effects beyond the two-particle
level.

Finally an intimate relation of the quantum potentials to density functional
theory has been explored which allows for DFT calculations without the need
to solve the Kohn-Sham equations.

\section{Acknowledgments}

We acknowledge stimulating discussions with V.~Filinov, W.D.~Kraeft,
D.~Kremp and M.~Schlanges. M.B. gratefully acknowledges hospitality
of the Physics Department of the University of Florida.

This work has been supported by the Deutsche
Forschungsgemeinschaft (BO-1366/2), the National Science Foundation
and the Department of Energy (grants DE FG03-98DP00218 and DE FG02ER54677),
as well as by grants for CPU time at the NIC J\"{u}lich and the Rostock Linux-Cluster
``Fermion''.


\begin{thebibliography}{10}

\bibitem{filinov-etal.jpa03ik} A.~Filinov, M.~Bonitz, and W.~Ebeling,
J. Phys. A: Math.Gen. {\bf 36},  5957 (2003).

\bibitem{boston97}\emph{Strongly Coupled Coulomb Systems}, G.~Kalman (ed.), Pergamon
Press 1998.

\bibitem{binz96}\emph{Proceedings of the International Conference on Strongly Coupled
Plasmas}, W.D.~Kraeft and M.~Schlanges (eds.), World Scientific,
Singapore 1996.

\bibitem{green-book}W.D.~Kraeft, D.~Kremp, W.~Ebeling, and G.~R\"{o}pke, \emph{Quantum
Statistics of Charged Particle Systems}, Akademie-Verlag Berlin
1986.

\bibitem{bonitz-book}M.~Bonitz, {}``Quantum Kinetic Theory'', B.G. Teubner, Stuttgart/Leipzig
1998.

\bibitem{Haberland}H. Haberland, M. Schlanges, W. Ebeling (eds.): Proc. 10th Int. Workshop
on the Physics of Nonideal Plasmas, Contrib. Plasma Phys. \textbf{41},
No 2-3 (2001).

\bibitem{kbt99}\emph{{}``Progress in Nonequilibrium Greens Functions''}, M.~Bonitz
(ed.), World Scientific Publ., Singapore 2000.

\bibitem{dasilva-etal.97}L.B.~Da~Silva et al., Phys. Rev. Lett. \textbf{78}, 483 (1997).

\bibitem{weir-etal.96}S.T.~Weir, A.C. Mitchell, and W.J.~Nellis, Phys. Rev. Lett. \textbf{76},
1860 (1996).

\bibitem{NormanStarostin}G.E.~Norman, and A.N.~Starostin, Teplofiz. Vys. Temp. \textbf{6},
410 (1968); \textbf{8}, 413 (1970), {[}Sov. Phys. High Temp. \textbf{6},
394 (1968); \textbf{8}, 381 (1970){]}
\bibitem{kremp}P.~Haronska, D.~Kremp, and M.~Schlanges, Wiss. Z. Universit\"{a}t
Rostock \textbf{98}, 1 (1987).
\bibitem{saumon}D.~Saumon, and G.~Chabrier, Phys. Rev. A \textbf{44}, 5122
(1991).
\bibitem{schlanges}M.~Schlanges, M.~Bonitz, and A.~Tschttschjan, Contrib. Plasma Phys.
\textbf{35}, 109 (1995).
\bibitem{BeEb99}D.~Beule et al., Phys. Rev. B \textbf{59}, 14177 (1999 ); Contrib.
Plasma Phys. \textbf{39}, 21 (1999)
\bibitem{fil_etal}V.S.~Filinov, V.E.~Fortov, M.~Bonitz, and P.R.~Levashov, JETP
Lett. \textbf{74}, 384 (2001) {[}Pis`ma v ZhETF \textbf{74}, 422
(2001){]}.
\bibitem{red-book}W.~Ebeling, W.D.~Kraeft, and D.~Kremp, ``Theory of bound states
and ionization equilibrium in plasmas and solids''.
Akademie-Verlag Berlin 1976; Russ. Transl. Mir Moscow 1979.

\bibitem{Margo} W.R.~Magro, D.M.~Ceperley, C.~Pierleoni and
B.~Bernu, Phys. Rev. Lett. {\bf 76}, 1240 (1996).

\bibitem{Militzer} B. Militzer and D.M.~Ceperley, Phys. Rev. E {\bf 63}, 066404
(2001).

\bibitem{trigger} S.A.~Trigger, W.~Ebeling, V.S.~Filinov, V.E.~Fortov, M.~Bonitz,
 JETP {\bf 96}, 465 (2003).

\bibitem{filinov2003} V.S. Filinov, M. Bonitz, W. Ebeling, and V.E. Fortov,
Plasma Physics and Controlled Fusion {\bf 43}, 743 (2001).

\bibitem{KTR94}D.~Klakow, C.~Toepffer, and P.-G. Reinhard, Phys. Lett. A \textbf{192},
55 (1994); J. Chem. Phys. \textbf{101}, 10766 (1994).

\bibitem{vova01}V.~Golubnychiy, M.~Bonitz, D.~Kremp, and M.~Schlanges, Phys. Rev.
E \textbf{64}, 016409 (2001).

\bibitem{qmd}For completeness we mention interesting concepts of quantum MD, such
as wave packet MD, e.g. \cite{KTR94} and Wigner function MD \cite{filinov-etal.02prb},
which are, however, beyond the scope of this paper.

\bibitem{filinov-etal.02prb}V.S.~Filinov, P.~Thomas, I.~Varga, T.~Meier, M.~Bonitz, V.E.~Fortov,
and S.W.~Koch,Phys. Rev. B \textbf{65}, 165124 (2002).

\bibitem{Ke63}G.~Kelbg, Ann. Physik, \textbf{12}, 219 (1963); \textbf{13}, 354;
\textbf{14}, 394 (1964).

\bibitem{Deu}
C.~Deutsch, Phys. Lett. {\bf 60}A, 317 (1977).

\bibitem{Rogers}
F.J.~Rogers, Phys. Rev. A {23}, 1008 (1981).

\bibitem{Deu2}
M.-M.~Gombert, H.~ Minoo, and C.~Deutsch, Phys. Rev. A {29}, 940
(1984).

\bibitem{perrot}
F.~Perrot and M.W.C.~Dharma-wardana, Phys. Rev. B {\bf 62}, 16536
(2000).

\bibitem{gombert}M.-M.~Gombert, H.~Minoo, Contrib. Plasma Phys. \textbf{29}, 355
(1989).

\bibitem{wagenknecht01}H.~Wagenknecht, W.~Ebeling, and A.~F\"{o}rster, Contrib. Plasma
Phys. \textbf{41}, 15 (2001).

\bibitem{storer}R.G.~Storer, J. Math. Phys. \textbf{9}, 964 (1968); A.D.~Klemm,
and R.G.~Storer, Aust. J. Phys. \textbf{26}, 43 (1973).


\bibitem{barker}A.A.~Barker, J. Chem. Phys. \textbf{55}, 1751
(1971).

\bibitem{Davis} B.~Devies and R.G.~Storer, Phys. Rev. {\bf 171},
150 (1968).

\bibitem{rohde} K.~Rodhe, G.~Kelbg, and W.~Ebeling, Ann. Phys. {\bf
22}, 1 (1968).

\bibitem{gombert2}M.-M.~Gombert, Phys. Rev. E \textbf{66}, 066407
(2002).

\bibitem{ceperley95rmp}D.M.~Ceperley, Rev. Mod. Phys. \textbf{65}, 279
(1995).

\bibitem{Kleinert}H.~Kleinert, \emph{Path Integrals in Quantum Mechanics, Statistics
and Polymer Physics}, World Scientific, Second edition, 1995.

\bibitem{Kleinert2}H.~Kleinert, Phys. Rev. D \textbf{57}, 2264
(1998).

\bibitem{spinflip} Since, in this case, individual electrons do not change their spin
projection, no exchange processes during collisions occur. This allows us to treat
the electrons with opposite spin projection as distinguishable
particles and to neglect exchange contributions to the pair potentials. In contrast,
in the scattering of two electrons with the same spin projection exchange does take
place which is taken into account by the potential (\ref{sing_t}). For a discussion of
this question, see C. Cohen-Tannoudji, B. Diu, and F. Laloe,
{\em Quantum Mechanics, vol. 2}, Ch. 14, John Wiley \& Sons 1977.

\bibitem{MilitzerPhD}B.~Militzer, PhD thesis, University Illinois,
(2000).

\bibitem{vova_phd} More details of the MD simulations are given in
V.~Golubnychiy, PhD thesis, Rostock University 2004.


%\bibitem{ebeling}W.~Ebeling, Ann. Physik (Leipzig) \textbf{21}, 315 (1968); \textbf{22},
%33, 383, 392 (1969); Physica \textbf{38}, 378 (1968); \textbf{40},
%290 (1968).

%\bibitem{EbFo91}W.~Ebeling, A.~F\"{o}rster, V.~Fortov, V.~Gryaznov, and A.~Polishchuk,
%\emph{Thermophysical properties of hot dense plasmas}, Teubner,
%Stuttgart-Leipzig 1991.

%\bibitem{Xu98}H. Xu, and J.\,P. Hansen, Phys. Rev. E \textbf{57}, 211 (1998).

%\bibitem{kelbg}W.~Ebeling, H.J.~Hoffmann, and G.~Kelbg, Contr. Plasma Phys. \textbf{7},
%233 (1967) and references therein.

\bibitem{mermin} N.D~Mermin, Phys. Rev. {\bf 137}, A1441 (1965).

\bibitem{zogaib} L.~Zogaib and J.~Dufty (to be published).

\bibitem{ks} W.~Kohn and L.~Sham, Phys. Rev. {\bf 140}, A1133 (1965).

\bibitem{dufty} J.W.~Dufty, unpublished.

%\bibitem{gee} The analytical from of the Kelbg potential is not well
%suited to describe the interaction of electrons with parallel
%spin. Instead, a suitable analytical approximation is given by
%Eq.~(\ref{u_e_d}).



%\bibitem{} F. Perrot, Phys. Rev. {\bf A20}, 586 (1979).

\end{thebibliography}
\end{document}